\documentclass[aps,pra,twocolumn]{revtex4-1}

\usepackage{amssymb}
\usepackage{color,graphicx}
\usepackage{amsmath}
\usepackage{amsbsy}
\usepackage{amsthm}
\usepackage{bbm}
\usepackage{bm}
\usepackage{epsfig}
\usepackage{times}
\usepackage{float}
\usepackage{subfigure}

\DeclareGraphicsExtensions{.pdf}

\newcommand{\nbar}{\overline{n}}
\newcommand{\ket}[1]{\left\vert {#1} \right\rangle} 
\newcommand{\bra}[1]{\left\langle {#1} \right\vert} 
\newcommand{\Tr}[2]{\mathrm{Tr}_{#1}\left[ {#2} \right]}

\begin{document}

\title{Quantum-limited estimation of continuous spontaneous localization}
\author{S. McMillen$^1$, M. Brunelli$^1$, M. Carlesso$^{2,3}$, A. Bassi$^{2,3}$, H. Ulbricht$^4$, M. G. A. Paris$^5$, and M. Paternostro$^1$}
\affiliation{$^1$Centre for Theoretical Atomic, Molecular and Optical Physics,
School of Mathematics and Physics, Queen's University, Belfast BT7 1nn, United Kingdom\\
$^2$Department of Physics, University of Trieste, Strada Costiera 11, 34151 Trieste, Italy\\
$^3$Istituto nazionale di Fisica nucleare, Trieste Section, Via Valerio 2, 34127 Trieste, Italy\\
$^4$Department of Physics and Astronomy, University of Southampton, SO17 1BJ, United Kingdom\\
$^5$Quantum Technology Lab, Dipartimento di Fisica, Universita` degli Studi di Milano, I-20133 Milano, Italy and InFn, Sezione di Milano, I-20133 Milano, Italy}
\date{\today}

\begin{abstract}
We apply the formalism of quantum estimation theory to extract information about potential collapse mechanisms of the continuous spontaneous localisation (CSL) form. 
In order to estimate the strength with which the field responsible for the CSL mechanism couples to massive systems, we consider the optomechanical interaction 
between a mechanical resonator and a cavity field. Our estimation strategy passes through the probing of either the state of the oscillator or that of the electromagnetic field that drives its motion. In particular, we concentrate on all-optical measurements, such as homodyne and heterodyne measurements. 
We also compare the performances of such strategies with those of  a spin-assisted optomechanical system, where the estimation of the CSL parameter is performed 
through time-gated spin-like measurements. 
\end{abstract}
\maketitle

Understanding the nature of the quantum-to-classical (QtC) transition is a long-sought problem that attracts an ever-growing attention~\cite{Leggett02, Adler04, AdlerBassi09, Weinberg12, Penrose96, Zurek}. While quantum mechanics has undergone exhaustive and {extremely} successful testings in the microscopic realm, the apparent absence of quantum manifestations at the macroscopic scale {cries for a deeper understanding}. {In particular, this lack of evidence reinforces the need of} assessing the causes for the emergence of classical mechanics from fundamental quantum evolution. The most widely accepted theory behind such process is {\it quantum decoherence}~\cite{Zurek}:  The environment surrounding any quantum system monitors its state continuously, practically collapsing the system's wavefunction and curtailing any quantum behaviour. Such process is conjectured to occur more quickly with the growing {\it size} of the system at hand. Under this regime, macroscopic superpositions would be possible in macroscopic systems perfectly isolated from their environment, a condition that is, for all practical purposes, not realisable.

However, a set of theories, usually referred to as collapse models (CMs), suggests an alternative route to the explanation of the QtC transition by putting forward fundamental underly- ing mechanisms responsible for the collapse of the wavefunction~\cite{BassiRMP}. The strength of this effect should increase with the size (mass) of the system, leaving microscopic (macroscopic) systems fully within the quantum (classical) realm. The key difference between CMs and standard quantum mechanics is that in the framework entailed by the former, perfectly isolated macroscopic objects would continue to act classically. 

Among the proposals put forward so far to test (or rule out) some of the currently formulated CMs~\cite{tests,tests2,tests3}, those based on the experimental platform of cavity optomechanics offer features of undemanding scalability of the mass of the system to be probed and high-sensitivity of measurement. Most remarkably, at variance with standardly pursued approaches~\cite{Marshall}, they bypass the need for the construction and quantum-limited management of large interferometers~\cite{Optoproposals,Optoproposals2}. notwithstanding such promising features, the investigation of CMs still poses considerable experimental challenges, and a winning strategy to their inference has not yet been singled out~\cite{McMillen}.

In this paper we propose that a potentially significant {\it boost} to the experimental assessment of CMs through optomechanical settings can come from the application of refined quantum inference techniques~\cite{estimation,estimationMatteo,estimationaltri} that have been so far successfully applied to achieve quantum-limited estimation of parameters of difficult accessibility in sophisticated quantum optics experiments~\cite{Berni,Berni2}. In order to fix the ideas and illustrate the pillars of our proposal in a concrete and relevant case, we focus on the continuous spontaneous localisation (CSL) model~\cite{GhirardiPearleRimini90,GhirardiGrassiBenatti95}, which is one of the simplest and most studied CMs~\cite{BassiRMP}. By applying the tools of quantum estimation theory to a paradigmatic cavity optomechanics system, we derive the ultimate bounds on the estimation precision of the core parameter entering the CSL model, thus going significantly beyond the achievements of any previous proposal in this context~\cite{Optoproposals,Optoproposals2}. Moreover, we identify a feasible, non-disruptive all-optical measurement strategy able to provide significant information on a CSL-affected nano-mechanical oscillator. Finally, we upgrade our system to a setup of hybrid cavity optomechanics which also includes a two-level system effectively coupled to the mechanical oscillator. By delegating the inference to measurements performed on the two-level system, we identify optimal instants of time and operating conditions that maximize the amount of information that could be extracted from the system itself. 
\par
The remainder of this work is organized as follows: In Sec.~\ref{s:EstTheory} we introduce the rudiments of quantum estimation theory, and present the main formal tools of the analysis that will be performed in the rest of the paper. In Sec.~\ref{s:Result} the action of the CSL model on a cavity optomechanical system is illustrated and  the main results are presented. In particular, in Sec.~\ref{s:OpticalEst} we address in detail the estimation via feasible measurements performed on the optical field, while in Sec.~\ref{s:SqueezingEst} we rule out the possibility of an enhancement in the estimation by resorting to squeezed-assisted strategies. In Sec.~\ref{s:HybridEst}
we address the estimability of the CSL parameter in a hybrid architecture featuring a coupling with a two-level system. Finally, in Sec.~\ref{s:Conclusions} we draw some conclusive remarks on our work.

\section{Elements of estimation theory}\label{s:EstTheory} 

Estimation theory is concerned with the inference of the parameters of a system based on a set of measured data. Quantum estimation theory studies the limitations to such inference due to quantum mechanics. In classical estimation theory, the Fisher information $\mathcal{I}_C(\Lambda)$ provides the amount of information about a parameter $\Lambda$ that is obtained from a particular measurement strategy. If the estimation is unbiased and based on $n$ of such measurements, then the uncertainty $\text{var}(\Lambda)$ associated with the estimation of the parameter in question is bounded by the Cramer-R{\'a}o bound $\textrm{var}(\Lambda) \geq [n\mathcal{I}_C(\Lambda)]^{-1}$. The quantity appearing on the right-hand side is the Fisher information of the parameter $\Lambda$, which is defined as 
\begin{equation}\label{FI}
\mathcal{I}_C(\Lambda)=\int {\left[\partial_\Lambda \ln p(x|\Lambda)\right]^2}p(x|\Lambda)\textrm{d}x,
\end{equation}
where $p(x|\Lambda)=\textrm{Tr}[\hat\rho(\Lambda) \hat{\text E}(x)]$ is the distribution of measurement outcomes $x$ conditional on the value of the parameter $\Lambda$ we wish to estimate, $\hat{\text E}(x)$ describes an element of the POVM linked to outcome $x$, and the integral spans all values of the measurement outcomes. 

In quantum estimation theory, the quantum Fisher information (QFI)  $\mathcal{I}_Q(\Lambda)$ concerns the information about $\Lambda$ contained in a quantum state $\hat\rho(\Lambda)$. It similarly satisfies a quantum Cramer-R{\'a}o bound $\textrm{var}(\Lambda)\ge[n\mathcal{I}_C(\Lambda)]^{-1} \geq [n\mathcal{I}_Q(\Lambda)]^{-1}$, and is given by
\begin{equation}\label{QFI}
\mathcal{I}_Q(\Lambda) = \textrm{tr}\left[\hat\rho(\Lambda)L^2(\Lambda)\right],
\end{equation}
where $L(\Lambda)$ is the symmetric logarithmic derivative (SLD), defined by $\partial_\Lambda{\hat\rho} = \left\{L(\Lambda),\hat\rho(\Lambda)\right\}/2$. The QFI is the optimized version of ${\cal I}_C(\Lambda)$ over all possible measurement strategies, which makes it explicitly independent of the specific measurement performed in order to infer $\Lambda$, and entails the ultimate bound set to the inference procedure by quantum mechanics, at least when one assumes that the measurement strategy does not depend explicitly on the parameter to be estimated \cite{niuCR}. 
The variance is only relevant to the precision of a measurement inasmuch as it relates to the mean value of the parameter. An alternative figure of merit for precision is the signal-to-noise ratio $S(\Lambda)=\Lambda^2/\textrm{var}(\Lambda)$. Using the quantum Cramer-R{\'a}o bound, we can also set an upper bound $S(\Lambda) \leq S_Q(\Lambda) \equiv \Lambda^2 \mathcal{I}_Q(\Lambda)$ which we call the quantum signal-to-noise ratio.

As it will be clarified later on in this paper, the system and evolution that we are going to address are Gaussian in nature. Therefore, we restrict the evaluation of the QFI to such class of states~\cite{GaussianQET}. An $n$-mode Gaussian state $\hat\rho(\Lambda)$ can be fully described by its $2n\times 2n$ covariance matrix $\sigma(\Lambda)$ with elements $\sigma_{ij}=\langle \{ \hat R_i, \hat R_j \} \rangle/2 - \xi_i \xi_j $ with $\xi_i = \langle \hat R_i \rangle$, $\hat {\bm R}$ 
the vector of canonical position and momentum operators, and where the average is calculated over the state of the system.

Moreover, when evaluating the Fisher information, we shall restrict our attention to local Gaussian measurements. One of such measurements can be formally described by a POVM whose elements are pure single-mode Gaussian states with covariance matrix  {$\sigma_\text{meas}=R\,\text{diag}(l/2,l^{-1}/2) R^T$}
which are then displaced to the point $x=(q,p)^T$ in the phase space. Here $l\in[0,\infty]$ parameterizes the degree of squeezing of the elements of the POVM, while $R=\cos\theta\openone-i\sin\theta\sigma_y$ is a rotation matrix (with $\sigma_y$ the usual $y$-Pauli matrix). If such a measurement is performed, $p(x|\Lambda)$ is then given by the Gaussian distribution $p(x|\Lambda)=\frac{\exp[-(x^T\sigma_p^{-1}x)/2]}{2\pi\sqrt{\det{\sigma_p}}}$ with covariance matrix $\sigma_p=\sigma(\Lambda) + \sigma_\text{meas}$.
This gives us 
\begin{equation}
\mathcal{I}_C(\Lambda)=\int^\infty_\infty \textrm{d}x \frac{e^{-\frac{1}{2}x^T\sigma_p^{-1}x}}{8\pi\sqrt{\det{\sigma_p}}}\eta(\Lambda,x)^2
\end{equation}
with $\eta(\Lambda,x)=x^T\sigma_p^{-1}(\partial_\Lambda\sigma_p)\sigma_p^{-1}x-\partial_\Lambda(\ln\det\sigma_p)$. Using standard Gaussian integration, we find the explicit form of the Fisher information for our model
\begin{equation}
\label{FI2}
\mathcal{I}_C(\Lambda)=
\frac{1}{2}\text{tr}{[(\sigma_p^{-1}\partial_\Lambda\sigma_p)^2]},
\end{equation}
where we have used Jacobi's theorem for the derivative of the determinant of a matrix to get the last expression. This result holds for Gaussian measurements of states with any kind of dependence on $\Lambda$. 

We now pass to the evaluation of the QFI for general $n$-mode Gaussian states.  To this aim, we introduce the symplectic matrix $\Omega=\bigoplus^n i \sigma_y$, which is instrumental to find the following expression for the SLD  
$L(\Lambda) = \hat R^T {\Phi} \hat R + \hat R^T {\bm \zeta} - \nu$,
whose quadratic dependence on $\hat R$ reflects the Gaussianity. Here, $\Phi$ is a $2n\times 2n$ real symmetric matrix, ${\bm \zeta}= \Omega^T\Sigma^{-1}(\partial_\Lambda{\bm \xi})$ is a real vector, and $\nu= \textrm{tr}(\Omega^T \sigma \Omega \Phi)$ is a scalar. Moreover, we have $\partial_\Lambda{\sigma} =2 \sigma\Omega \Phi \Omega^T \sigma - \frac{\Phi}{2}$. The QFI can be expressed in terms of these quantities: $\mathcal{I}_Q(\Lambda) = \textrm{tr}[\Omega^T\partial_\Lambda\sigma\Omega{\Phi}]+\partial_\Lambda{\bm\xi}^T\sigma^{-1}\partial_\Lambda{\bm\xi}$

Following the procedure illustrated in the Appendix, we can determine the explicit form of $\Phi$,
and thus a compact expression for the QFI of a single mode in terms of $\sigma$ and its derivative with respect to the parameter to estimate:
\begin{equation}\label{QFI2}
\mathcal{I}_Q(\Lambda)=\frac{\det{(\partial_\Lambda\sigma)}^2\textrm{tr}{[((\partial_\Lambda\sigma)^{-1}\sigma)^2]}+\frac{1}{2}\det{(\partial_\Lambda\sigma)}}{2\det{\sigma}^2-\frac{1}{8}}.
\end{equation}
Eqs.~\eqref{FI2} and \eqref{QFI2} embody the main tools of our analysis, which will address the covariance matrix of the CSL-affected optomechanical system illustrated in the next Section.

\section{The model and the core results}\label{s:Result}
The CSL model modifies the standard Schr\"odinger equation by adding nonlinear stochastic terms. The model makes use of two parameters, $\gamma$ and $r_c$ which will be introduced shortly. It can be shown explicitly that for the optomechanical system we will be working with the evolution is conveniently described by the following linear evolution:
\begin{equation}\label{numerouno}
i \hbar \frac{\mathrm{d}}{\mathrm{d}t} \left| \Psi_t (q) \right\rangle = \left(\hat{H}_0 + \hat{V}_t \right)\left| \Psi_t (q) \right\rangle
\end{equation}
with the potential $\hat{V}_t = -\hbar \sqrt{\eta} \hat{q} w_t$, where $w_t$ describes white noise with $\langle w_t\rangle = 0$ and $\langle w_t w_{t'}\rangle = \delta(t-t')$, and 
\begin{equation}\label{Eta}
\eta = \frac{\gamma}{3 m_0^2} \sum_{k=1}^3 \int \frac{e^{-\frac{|{\bf r} - {\bf r}'|^2}{4 r_c^2}}}{\left(2 \sqrt{\pi} r_c \right)^3} \partial_{r_k} \rho_d({\bf r}) \partial_{r_k'} \rho_d({\bf r}') \mathrm{d}{\bf r}\mathrm{d}{\bf r}'.
\end{equation}
where $m_0 = 1 \textrm{amu}$, $\rho_d({\bf r})$ is the mass density of the system subjected to the effects of the collapse mechanism and $r_c$ is a characteristic length-scale, typically assumed to be $~100$nm, above which reduction effects would be relevant. The crucial quantity appearing in Eq.~\eqref{Eta} is $\gamma$, which represents the coupling strength between the system and the collapse noise, and is the parameter we wish to estimate using the estimation theory illustrated above. 
Its actual value is the subject of uncertainties~\cite{BassiRMP}: Ghirardi, Pearle and Rimini suggested a value for $\gamma_{GRW} \simeq 10^{-36}\text{m}^3\text{s}^{-1}$~\cite{GhirardiPearleRimini90}, whereas Adler suggests $\gamma_A \simeq 10^{-28}\text{m}^3\text{s}^{-1}$~\cite{Adler2006}. In our analysis, we will be interested in exploring the implications that the different expected values of $\gamma$ have on the precision associated with a chosen strategy. Data coming from various experiments and performed at a broad range of energy scales tightly constrain the range of possible values for $r_c$, which makes it unnecessary to invoke quantum estimation theory methods to further assess the variability of such parameter~\cite{VinanteExclusion}.

The system we consider is an optomechanical cavity of length $L$ pumped externally with laser light of strength $\mathcal{E}$ and frequency $\omega_0$. 
The Hamiltonian of the system (in a rotating frame at the frequency of the external pump) reads~\cite{AspelmeyerRMP}
\begin{equation}\label{Hamiltonian0}
\begin{aligned}
\hat{H}_0 &= \frac{\hbar \Delta}{2}(\hat X^2+\hat Y^2)+ \frac{\hbar \omega_m}{2}(\hat{q}^2 + \hat{p}^2)- \frac{\hbar \chi_0}{2} (\hat X^2+\hat Y^2) \hat{q} \\
&{+ i \hbar \sqrt2 \mathcal{E}\hat Y}
\end{aligned}
\end{equation}
where we have rigidly shifted the energy of the system by $-\hbar\Delta/2$ and neglected a very small  frequency shift of the mechanical motion. In Eq.~\eqref{Hamiltonian0} $\hat{q}$ and $\hat{p}$ are dimensionless position and momentum operators for the mechanical oscillator of effective mass $m$ (oscillating at frequency $\omega_m$), $\hat{X}$ and $\hat{Y}$ are the quadrature operators for the cavity field, and $\Delta=(\omega_c - \omega_0)$ is the cavity-pump detuning. The third term in Eq.~\eqref{Hamiltonian0} describes the optomechanical interaction with coupling strength $\chi_0 = (\omega_c /L)\sqrt{\hbar/m \omega_m}$. The last term describes the coupling between the cavity and the (classical) driving field, $\mathcal{E}$ being the rate of pumping. 

The dynamics of this system has been studied extensively, and we refer to Ref.~\cite{Paternostro2006} for a detailed formal analysis. For the sake of our scopes, it is sufficient to mention here that, under the assumptions of strong external driving and high-quality mechanical motion, which is generally affected by incoherent Brownian noise at temperature $T$, the optomechanical evolution can be split into a (classical) mean-field part, and a (quantum) fluctuation-affected one.  
The latter is what we concentrate on, as it encompasses non-trivial correlations between the optical and mechanical sub-parts of our system~\cite{Paternostro2006}. We thus assume to be in a position to expand any operator $\hat O$ of the system as $\hat O=\overline{O}+\hat{\delta O}$, where $\overline{O}$ is the corresponding mean part, and define the vector of zero-mean fluctuations $\hat u=(\delta\hat{q},\delta\hat{p},\delta\hat{X},\delta\hat{Y})^T$, which we use in order to define the covariance matrix of the fluctuations $\sigma_f$ with elements $(\sigma_f)_{ij}=\langle\{\hat u_i,\hat u_j\}\rangle/2$, which can be shown to evolve according to the equation~\cite{Paternostro2006}
\begin{equation}
\label{tdLy}
\partial_t\sigma_f=A\sigma_f+\sigma_f A^T+D,
\end{equation}
where we have introduced the drift matrix $A$ and the noise one $D$ given by 
\begin{equation}\label{eq:A}
A =
\left[
\begin{matrix}
0 & \omega_m & 0 & 0 \\
-\omega_m & -\gamma_m & \chi & 0 \\
0 & 0 & -\kappa & \Delta \\
\chi & 0 & -\Delta & -\kappa
\end{matrix}\right],\,
D=
\left[
\begin{matrix}
0&0&0&0\\
0&\Gamma(\Lambda)&0&0\\
0&0&\kappa&0\\
0&0&0&\kappa
\end{matrix}
\right].
\end{equation}
In these expressions, $\gamma_m$ is the natural damping rate of the mechanical motion, $\kappa$ is the decay rate of the cavity field, $\chi = \sqrt{2}\chi_0 {\cal E}/{\sqrt{\kappa^2 + \Delta^2}}$ is an effective optomechanical coupling rate and $\Gamma(\Lambda)=\gamma_m(2\nbar+1)+\Lambda$ with $\Lambda =\hbar\eta/m\omega_m$, and $\nbar$ the mean number of thermal phonons in the initial state of the mechanical oscillator (which is assumed to be a Gibbs state at the environmental temperature $T$). Quite evidently, the CSL mechanism enters the dynamics of the optomechanical system only through the noise matrix $D$ and in the form of an additional source of mechanical damping. Alternatively, the CSL effect can be interpreted as an increased equilibrium temperature of the mechanical system (cf. Ref.~\cite{McMillen} and Bahrami {\it et al.} in~\cite{Optoproposals,Optoproposals2}) that changes $\nbar$ to 
\begin{equation}\label{equiv}
\nbar_{\text{csl}}=\nbar+\frac{\Lambda}{2\gamma_m}.
\end{equation}
Therefore, estimating $\Lambda$ is equivalent, from this viewpoint, to the estimation of the equilibrium temperature of the mechanical system~\cite{Optoproposals,Optoproposals2}. While the optimal estimation strategy for the inference of temperature of an equilibrium harmonic oscillator has been found to be provided by measurements of its energy (the QFI being proportional to the variance of the energy of the oscillator)~\cite{thermometry}, here we would like to exploit the coupling between the mechanical system and the cavity field to devise implementable strategies for the inference of $\Lambda$. The inspection of Eq.~\eqref{Hamiltonian0} shows that the latter is only coupled to the position of the mechanical oscillator, which would not be sufficient to infer its energy directly. We thus proceed to a full-fledged analysis of the results achievable through the use of quantum estimation theory in the context set by this paper.  

From the analysis reported in Sec.~\ref{s:EstTheory}, it is clear that we need to evaluated the covariance matrix $\sigma_f$. This can be done straightforwardly at the steady state, where $\sigma^{ss}_f(\Lambda)$ is the solution of the Lyapunov equation $A\sigma^{ss}_f(\Lambda)+\sigma^{ss}_f(\Lambda) A^T=-D$. The explicit form of such solution can be deduced from the expressions reported in~\cite{Paternostro2011} with the replacement $\nbar\to\nbar_{\text{csl}}$. It is sufficient to mention that $\sigma^{ss}_f(\Lambda)$ takes the general form
\begin{equation}
\sigma^{ss}_f(\Lambda)=
\begin{pmatrix}
\sigma_M & \sigma_C \\
\sigma_C^T & \sigma_L \\
\end{pmatrix}.
\end{equation}
Here $\sigma_M$ ($\sigma_L$) encompasses the covariances of the mechanical (optical) subsystem, while $\sigma_C$ brings about the optomechanical correlations. In general $\sigma_M$ turns out to be a diagonal matrix, while all the entries of $\sigma_L$ are in general non-null. Moreover, the dependence of $\sigma_M$ on the rescaled CSL parameter $\Lambda$ is found to be linear, and we can write
\begin{equation}
\sigma_M=
\begin{pmatrix}
\alpha_1 + \beta_1\Lambda & 0 \\
0 & \alpha_2 + \beta_2\Lambda
\end{pmatrix},
\label{mechcovmtrx}
\end{equation}
where $\alpha_{1,2}$ and $\beta_{1,2}$ are scalars whose explicit form can be found in Appendix~C. In fact, all the elements of $\sigma^{ss}_f(\Lambda)$ are linear functions of $\Lambda$. 

\begin{figure}[t]
{\bf (a)}\\
\includegraphics[width=0.75\columnwidth]{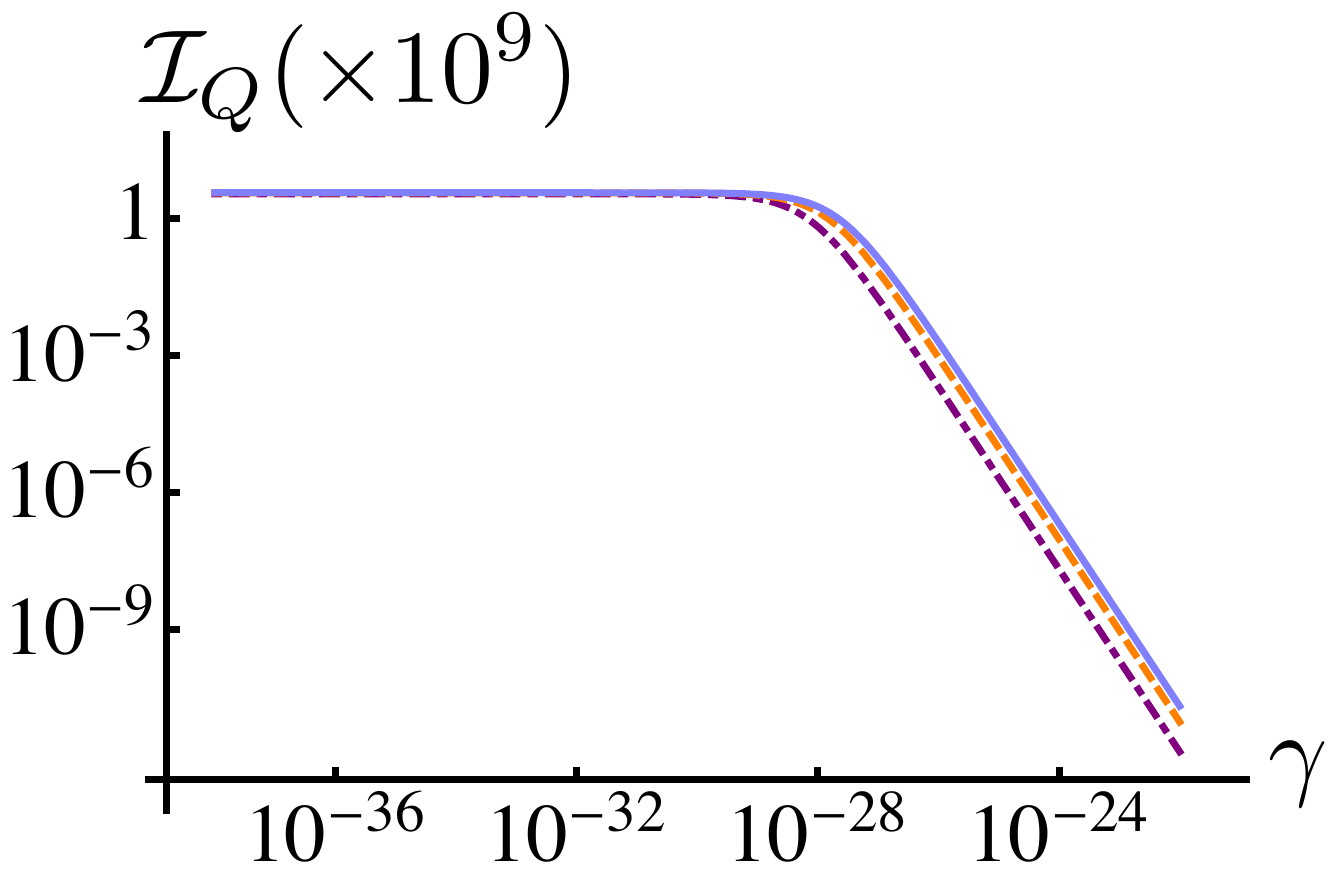}
\\
{\bf (b)}\\
\includegraphics[width=0.77\columnwidth]{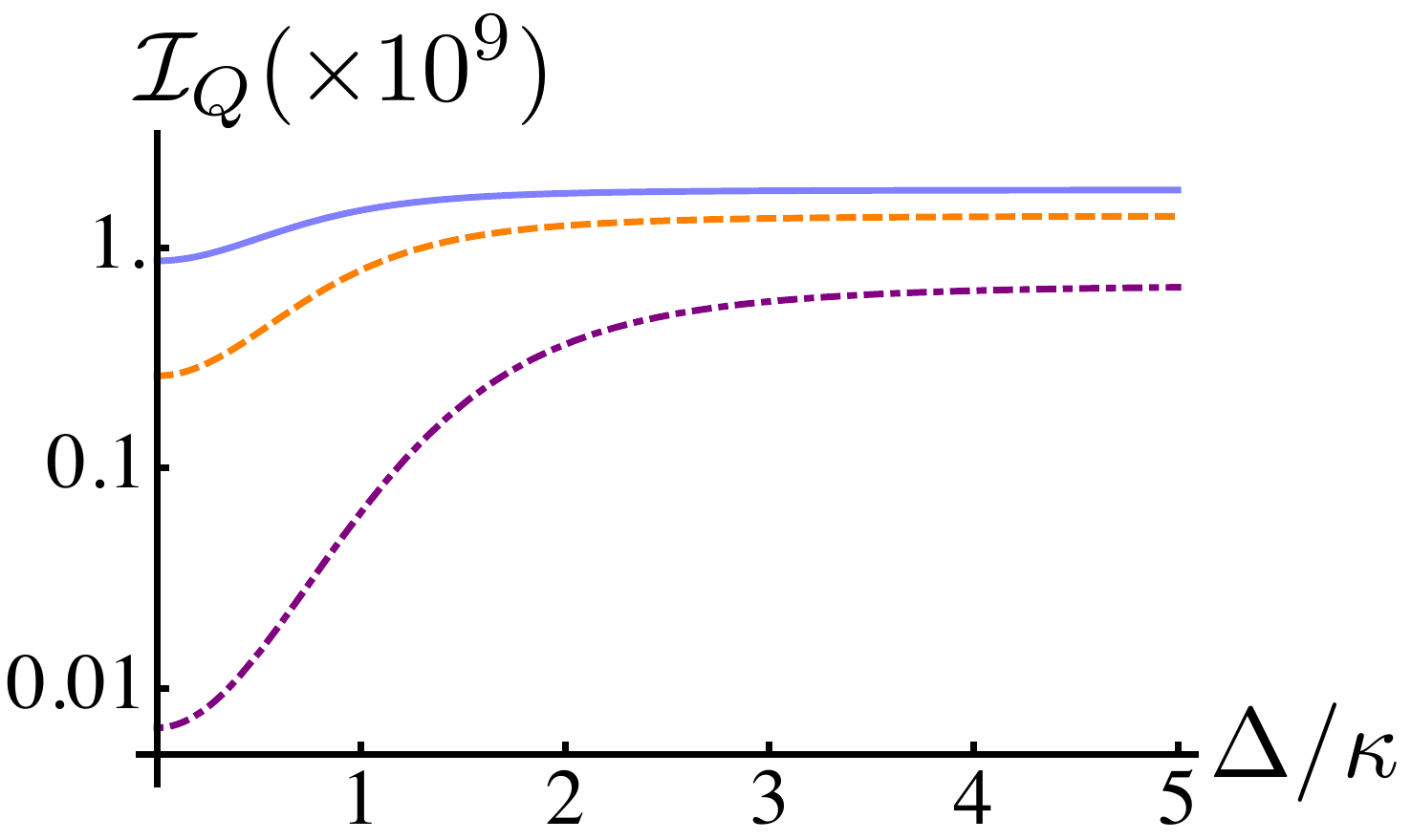}
\caption{(Color online) {\bf (a)} Logarithmic plot of the QFI ($\times10^{9}$) associated with the state of the mechanical system, plotted against the coupling strength $\gamma$ to the CSL noise field. We have used a mechanical oscillator of mass $15\textrm{ng}$ (dot-dashed magenta curve), $150$ng (dashed orange curve) and $500$ng (light blue, solid curve). Other parameters are $\omega_m/2\pi = 2.75 \times 10^5 \textrm{Hz}$, $\gamma_m/2\pi = \omega_m/10^5$, $L=25\textrm{mm}$, laser power  $\mathcal{P}=2\textrm{mW}$, $\kappa = 5 \times 10^7\textrm{Hz}$, $T=1\textrm{mK}$, $\Delta=5\kappa$. 
{\bf (b)} We plot ${\cal I}_Q$ against $\Delta/\kappa$ using the same parameters as in the main panel but for $\gamma=\gamma_A$. The three curves correspond to the values of mass used for the main panel.\label{FImauro} }
\end{figure}

Having the covariance matrix of the mechanical system at hand, we can apply the formalism of quantum estimation theory illustrated in the previous Section to find the analytic form of the QFI  
\begin{equation}\label{QFImechanics}
\begin{aligned}
\mathcal{I}_Q(\Lambda)&= 
\frac{4[\beta_1\beta_2+2\beta^2_2(\alpha_1+\beta_1\Lambda)^2+2\beta^2_1(\alpha_2+\beta_2\Lambda)^2]}{D(\Lambda)}
\end{aligned}
\end{equation}
with $D(\Lambda)=16 (\alpha_1+\beta_1 \Lambda )^2 (\alpha_2+\beta_2 \Lambda )^2-1$. We have replicated this result also using the fidelity based approach to calculating the QFI \cite{Pinel2013}. Eq.~\eqref{QFImechanics} is the basis of the study shown in Fig.~\ref{FImauro}, where we show the behavior of ${\cal I}_Q$ for the mechanical system within an ample range of values of the CSL parameter $\gamma$, including both the Ghirardi-Pearle-Rimini and the independent Adler's estimate. While the choice of parameter made in Fig.~\ref{FImauro} is not linked to a specific experimental implementation and has been dictated by the visibility of the curves, the displayed behavior should be considered as canonical. A few considerations are in order. First, the variance associated with even the best measurement strategy (as entailed by the QFI) is very large showing that, at least with this setup, the estimate of $\gamma$ would not be able to help ruling out the actual CSL influences. Second, the QFI appears to be largely insensitive of the actual value taken by the CSL coupling strength for values of $\gamma$ within a rather large range. In particular, the estimate $\gamma_{GPR}$ falls well within such an {\it insensitive} region. The behavior of the QFI changes, instead, dramatically for $\gamma\gtrsim\gamma_A$, showing a knee almost in correspondence of the estimate by Adler, and decreasing quickly as $\gamma$ increases. Finally, in Fig.~\ref{FImauro} {\bf (b)} we show that, at such {\it critical} value of the CSL coupling parameter, mechanical systems of a larger mass offer enhanced sensitivity and thus a lower variance associated with the estimation of $\gamma_A$, in accordance with the expectation that stronger reduction effects should be expected in massive systems. Moreover, as large values of $\Delta$ with respect to the cavity line width $\kappa$ correspond to weaker optomechanical couplings (cf. the form of parameter $\chi$), we conclude that weakly-perturbed mechanical oscillators offer better performances. 


\begin{figure*}[t]
{\bf (a)}\hskip4cm{\bf (b)}\hskip4cm{\bf (c)}\hskip4cm{\bf (d)}\\
\includegraphics[width=0.50
\columnwidth]{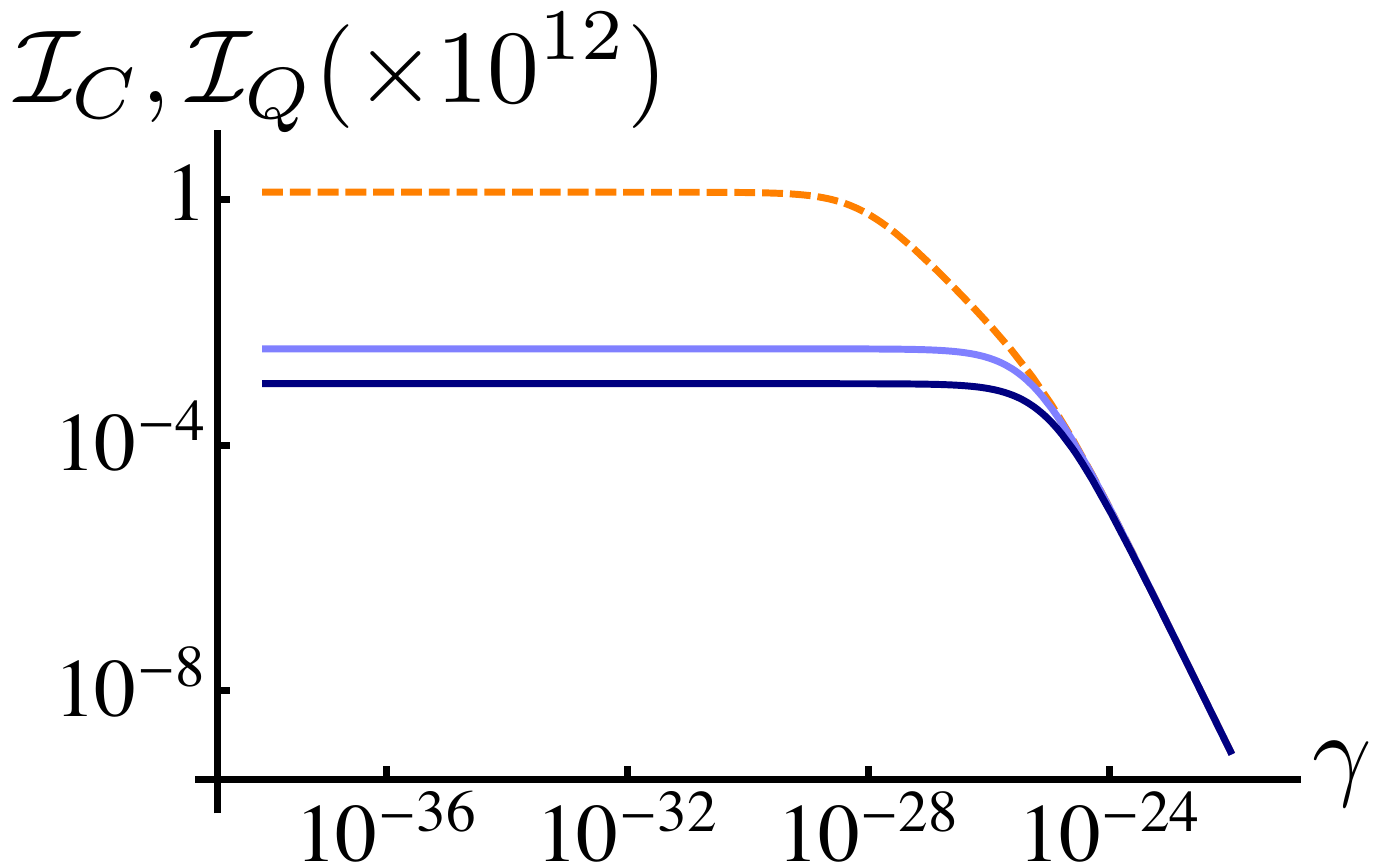}\,\,\includegraphics[width=0.60
\columnwidth]{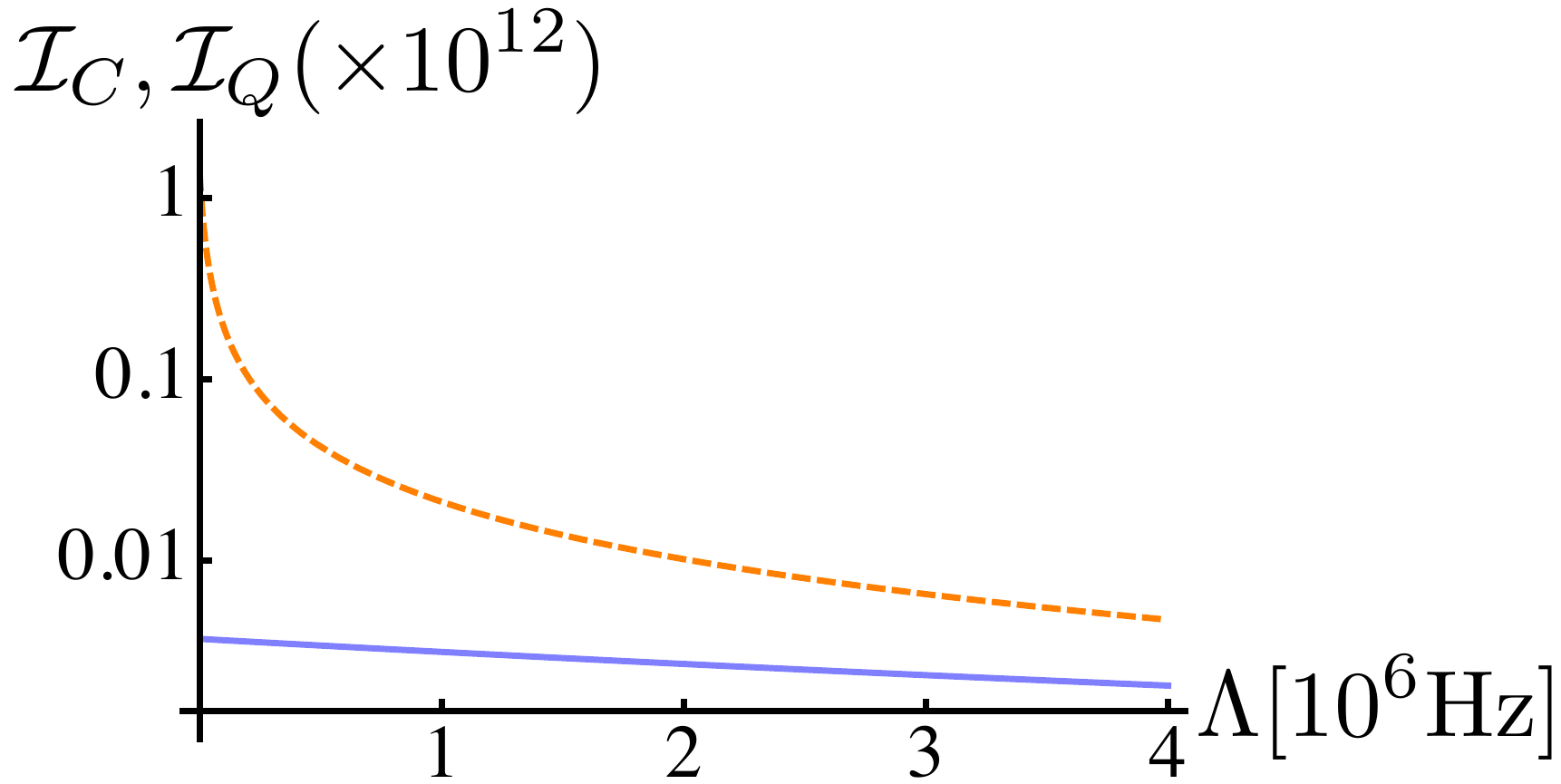}\,\,\includegraphics[width=0.50\columnwidth]{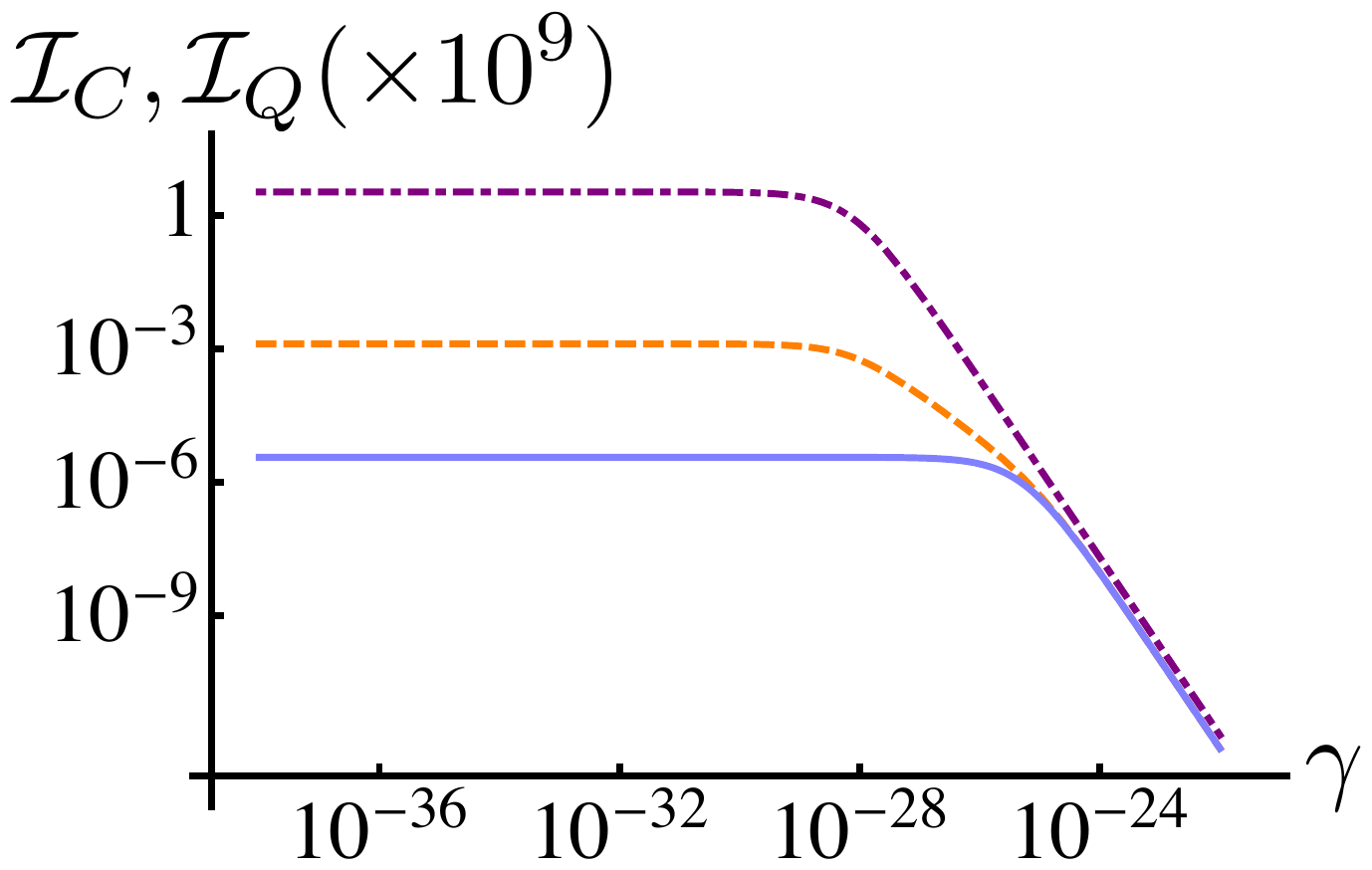}\,\,\includegraphics[width=0.50\columnwidth]{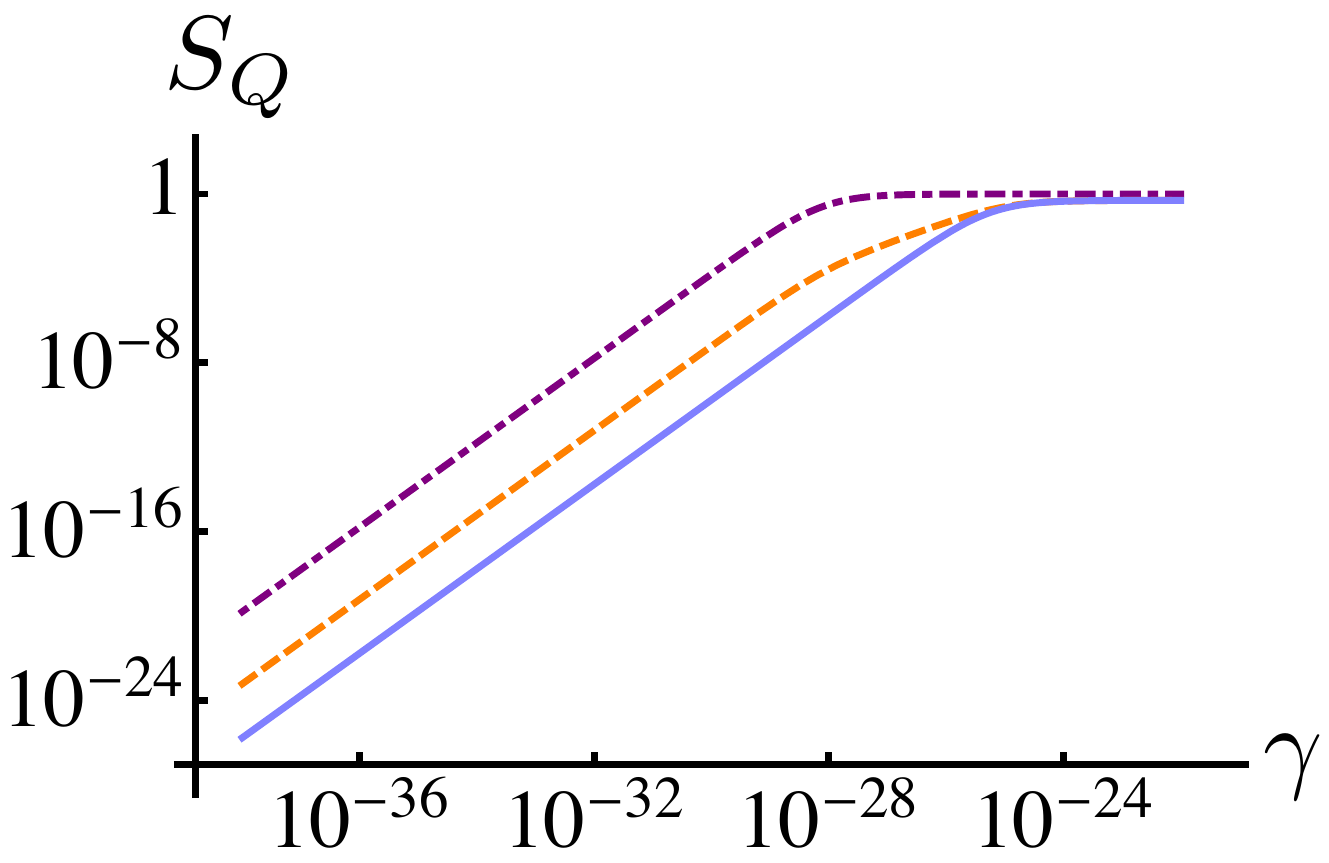}
\caption{(Color online) {\bf (a)} The QFI (orange dashed curve) and the Fisher information $\mathcal{I}_C$  (both $\times10^{12}$) associated with the optical subsystem probed by a homodyne (light blue curve) and a heterodyne measurement (dark blue curve) are plotted against the coupling strength $\gamma$ to the CSL noise field. 
{\bf (b)} QFI (orange dashed) and Fisher information $\mathcal{I}_C$ (blue) [both $\times10^{12}$] of the optical subsystem when performing homodyne detection ($l=0$) and choosing $\theta=0$, plotted against the rescaled CSL parameter $\Lambda$ (in units of $10^6$Hz).
{\bf (c)} We compare the QFI associated with the state of the mechanical subsystem (magenta, dot-dashed), the QFI of the state of the optical field (orange dashed), and the Fisher information resulting from the performance of a homodyne measurement on the optical field. All such quantities have been rescaled by $10^{9}$. 
{\bf (d)} We compare the quantum signal-to-noise ratio associated with the state of the mechanical subsystem (magenta, dot-dashed), the optical field (orange dashed), and the signal-to-noise ratio of a homodyne measurement on the optical field, plotted against the coupling strength $\gamma$ to the CSL noise field. 
All the panels refer to a value of the mass of $15\textrm{ng}$, while the other parameters are the same as in Fig.~\ref{FImauro}.}
\label{FI-QFIoptics}
\end{figure*}
\par

\subsection{Estimation through the optical subsystem}\label{s:OpticalEst}
Looking now for the sort of precision that an actual measurement strategy would be able to achieve when implemented on the mechanical system, it is important to stress the lack of direct access to the physical properties of a mechanical oscillator in an optomechanical cavity: in fact, the direct measurement of the mechanical oscillator is either considered undesirable due to the strong back-action entailed by direct probing, or technically challenging in light of the usual necessity of operating an optomechanical device at low pressure (which requires ultra-high vacuum chambers) and temperature (thus requiring a cryostat). 
We thus address the estimation of $\Lambda$ from a different perspective, and investigate the amount of information that can be extracted by performing local Gaussian measurements on the state of the optical subsystem instead. This approach is meaningful in light of the optomechanical coupling, which encodes information on the CSL-affected mechanical oscillator onto appropriate degrees of freedom of the cavity field, and indeed embodies the standard way of inferring information on the mechanical motion~\cite{AspelmeyerRMP,Paternostro2006}. 
We have thus calculated the QFI associated with the steady state of the optical field, repeating the analysis displayed in Fig.~\ref{FImauro}. A noticeable difference between the two cases is, though, that the optical covariance matrix $\sigma_L$ is, in general, non diagonal, which makes the provision of a fully analytical expression for ${\cal I}_Q$ inconvenient, in this case. nevertheless, it is possible to assed the QFI against $\gamma$, as shown in Fig.~\ref{FI-QFIoptics} {\bf (a)}, which displays similar features to those revealed when assessing the all-mechanical case (notice, though, the even smaller values taken by ${\cal I}_Q$, which is a clear result of the indirect probing that we are considering here).

Quite remarkably, we are now in a position to consider suitable all-optical measurement strategies. Rather than trying to identify the measurement that renders ${\cal I}_C={\cal I}_Q$, we decided to take a pragmatic approach based an only consider experimentally non-demanding measurements. Moreover, in order to make use of the powerful framework for Gaussian states probed by Gaussian measurements illustrated above, we shall restrict the class of probing strategies to local Gaussian ones, and consider both homodyne and heterodyne measurements. This can be done very conveniently using our parameterisation of $\sigma_\text{meas}$ and choosing appropriately the value of $l$. In fact, for $l=0$ or $\infty$ we would implement homodyne detection, while heterodyning would correspond to $l=1$. While the choice of $\theta$ would be inessential for the latter instance, the value of such angle determines the direction, in phase space, along which homodyning is performed. As one could expect, this is an important parameter in the determination of the best estimation strategy. 

We have calculated the Fisher information ${\cal I}_C$ in Eq.~\eqref{FI2} by varying the choice of $l$ and $\theta$ finding that, although none of the chosen strategies is {\it optimal} over the range of values of $\gamma$ up to the estimate given by Adler, homodyning appears to be superior to heterodyne measurements. This is illustrated in Fig.~\ref{FI-QFIoptics} {\bf (a)}, where one can appreciate that homodyne measurements result in values of the Fisher information that are about one order of magnitude larger than those corresponding to heterodyning for $\gamma$ up to $10^{-24}$m$^3$s$^{-1}$.
Interestingly, we find that while for $\gamma < \gamma_A$ the Fisher information and the QFI are both flat (thus implying no improvement in the precision of the estimation of $\gamma$ across tens of orders of magnitude), for $\gamma > \gamma_A$ the two figures of merit get very close to each other, regardless of the measurement strategy being implemented [cf. Fig.~\ref{FI-QFIoptics} {\bf (b)}, where for $\gamma=\gamma_A$ we have $\Lambda\simeq10^5$Hz]. The low values achieved by both ${\cal I}_Q$ and ${\cal I}_{C}$ for $\gamma > \gamma_A$, though, demonstrate the very weak sensitivity of the setting that we have chosen to reduction models characterized by coupling strengths in such region of values. In Fig.~\ref{FI-QFIoptics} {\bf (c)} we finally summarize our analysis so far by comparing the mechanical QFI, the optical one and the Fisher information associated with the homodyne probing of the optical field's state. 
In Fig.~\ref{FI-QFIoptics} {\bf (d)} we examine the quantum signal-to-noise ratio $S_Q$ against the CSL coupling strength. In the region where $\gamma > \gamma_A$, we see that $S_Q$ approaches $1$. This is identical to the behaviour of the quantum signal-to-noise ratio for estimation of the temperature of a single quantum harmonic oscillator in a thermal state. As shown in Eq (\ref{equiv}) the CSL mechanism behaves like an additional heating term, which dominates when $\Lambda \geq 2 \gamma_m \nbar$. For our choice of parameters this inequality is saturated  when $\gamma= 10^{-28.1}\approx \gamma_A$, thus for larger values of $\gamma$ the CSL noise term becomes the main contributor of heat to the system.

\begin{figure*}[t]
{\bf (a)}\hskip6cm{\bf (b)}\hskip6cm{\bf (c)}\\
\includegraphics[width=0.72\columnwidth]{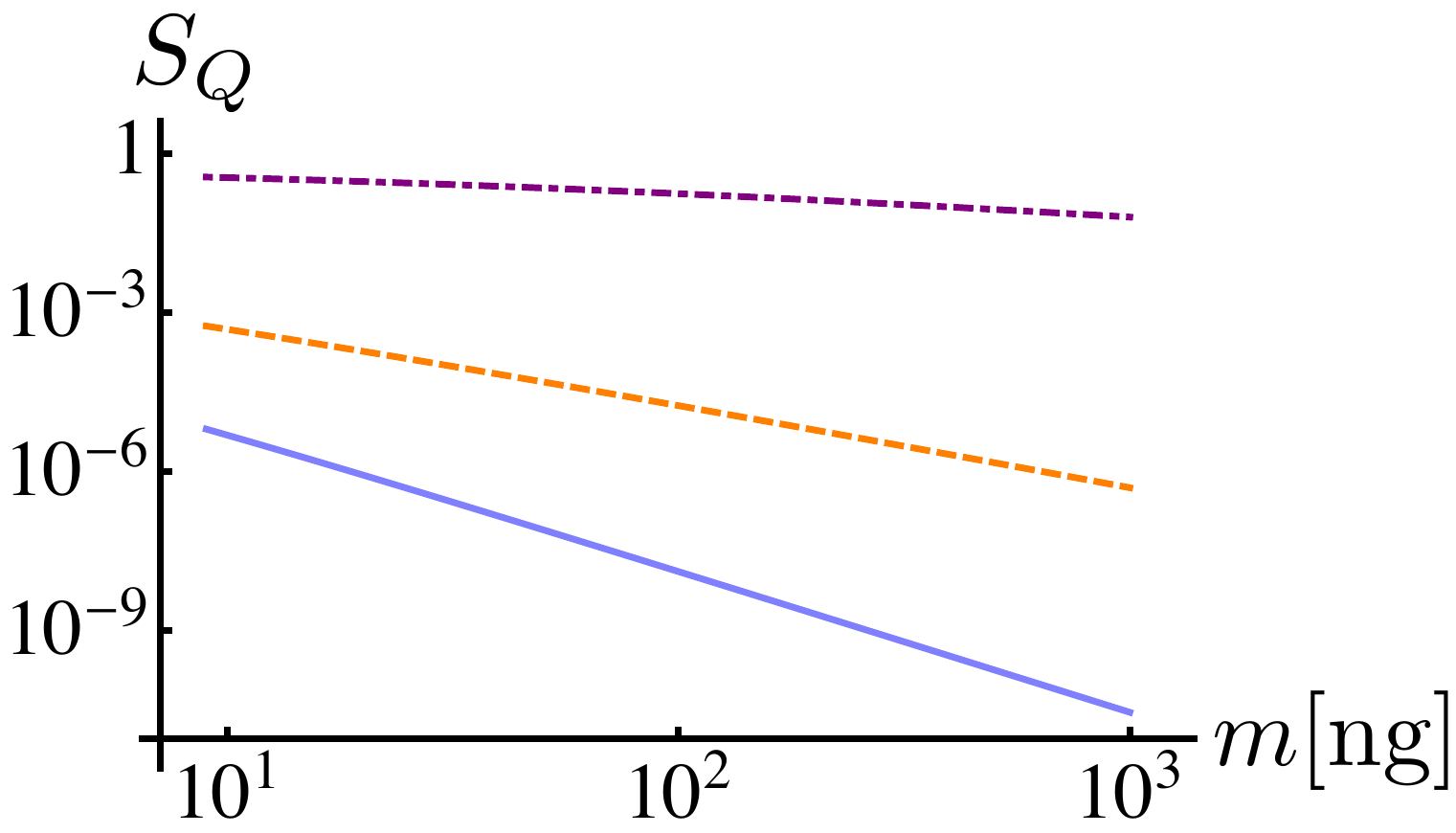}\,\,\includegraphics[width=0.70
\columnwidth]{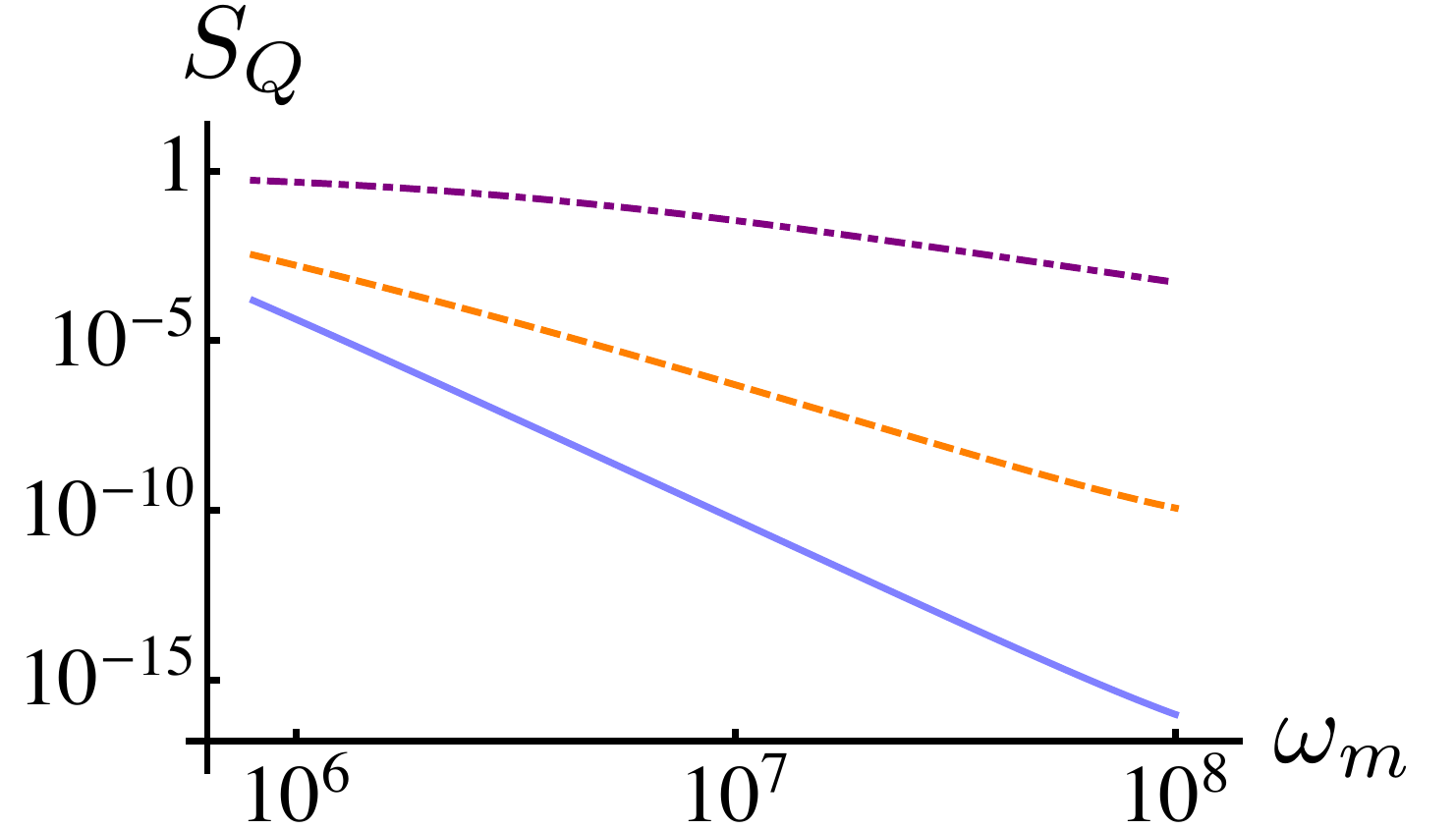}\,\,\includegraphics[width=0.67\columnwidth]{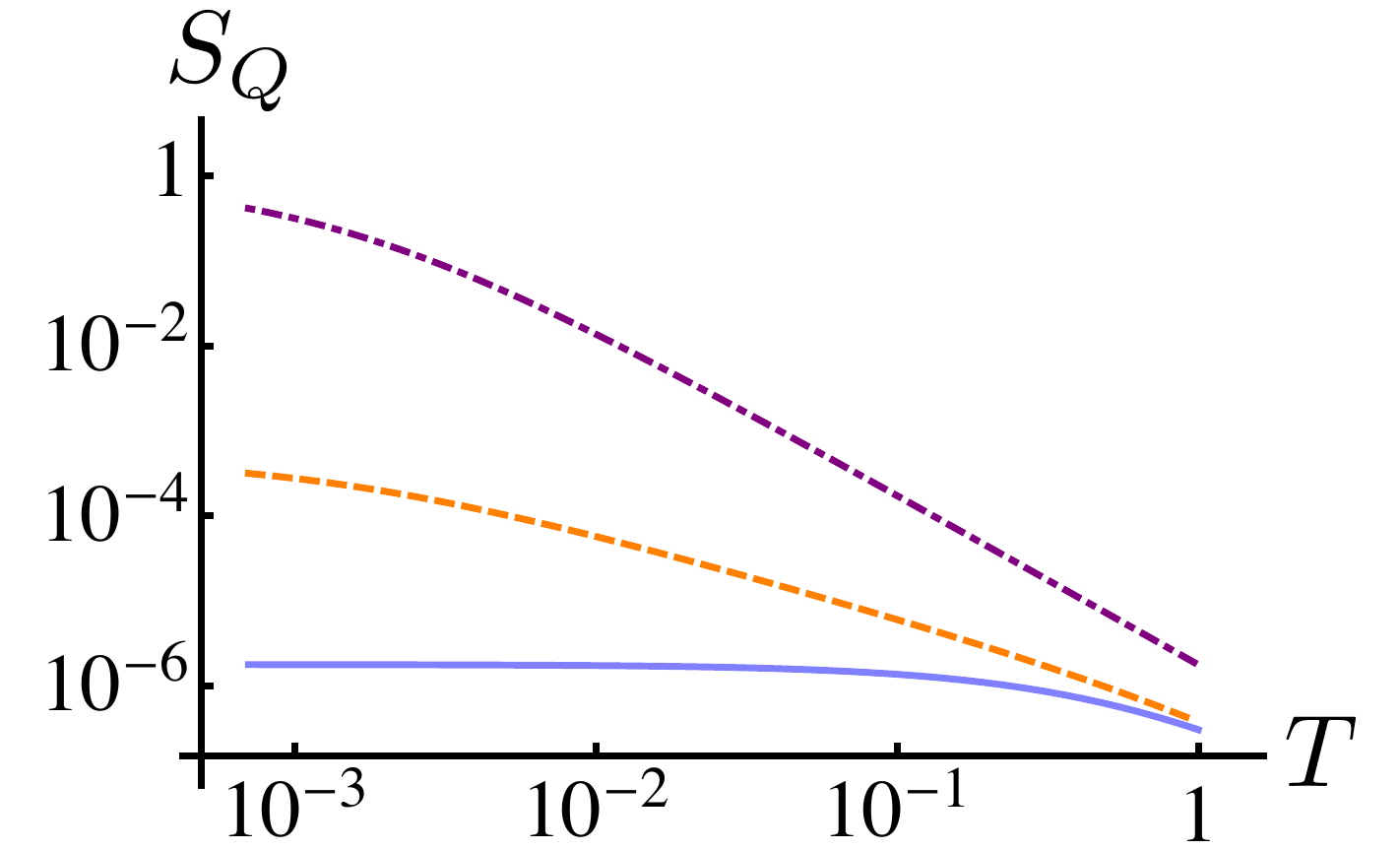}
\caption{(Color online) {\bf (a)} Logarithmic plot of the quantum signal-to-noise ratio associated with the QFI of the mechanical subsystem (magenta, dot-dashed), the optical subsystem (orange dashed curve) and the signal-to-noise associated with the Fisher information of a homodyne measurement on the optical subsystem (solid blue curve) plotted against the mass $m$ of the mechanical oscillator in nanograms. 
{\bf (b)} The same three quantities with the same colour designation, with the mass fixed at $15\textrm{ng}$, plotted against the frequency $\omega_m$ of the mechanical oscillator.
{\bf (c)} The same quantities plotted now against the initial temperature $T$ of the system. In all the panels we have used the same parameters as in Fig.~\ref{FImauro}.
}
\label{signaltonoiseparams}
\end{figure*}

In Fig.~\ref{signaltonoiseparams} the sensitivity of our system to some crucial parameters is examined by analysing the quantum signal-to-noise ratio $S_Q$ associated with the QFI of the mechanical and optical subsystems and the signal-to-noise ratio associated with the Fisher information of a homodyne measurement on the optical subsystem. First of all, and as already demonstrated in Fig.~\ref{FI-QFIoptics} {\bf (d)}, $S_Q$ for the mechanical subsystem is larger than $S_Q$ for the optical one, since the cavity field is only indirectly affected by the CSL mechanism. 
In Fig.~\ref{signaltonoiseparams} {\bf (a)} and {\bf (b)} the three quantities are shown to decrease when the mass of the oscillator or the mechanical frequency is increased. This results from the dependence of the optomechanical coupling $\chi_0$ on $(m\omega_m)^{-1/2}$. This is in agreement with the results of Nimmrichter et. al. in \cite{Optoproposals}. These figures also show that the gap widens between the quantum and classical signal-to-noise ratios as these quantities are increased, so the measurement procedure we describe moves further from optimality.  In Fig.~\ref{signaltonoiseparams} {\bf (c)} the three quantities decrease when the initial temperature $T$ of the system is increased. This is expected as a higher initial temperature implies a higher amount of thermal noise. We also see that the gap closes between the quantum and classical signal-to-noise ratios at around $1\textrm{K}$. The mechanical $S_Q$ is still higher than the optical one, as justified above, but the classical signal-to-noise ratio reaches $S_Q$ for the optical subsystem, saturating the Cramer-R\'{a}o bound. This can be explained by the fact that at such high temperatures, the system will behave classically, and therefore the limit on the precision set by quantum mechanics is equal to the classical one.

\subsection{Squeezing-assisted estimation}\label{s:SqueezingEst}

\begin{figure}[b]
\includegraphics[width=0.75\columnwidth]{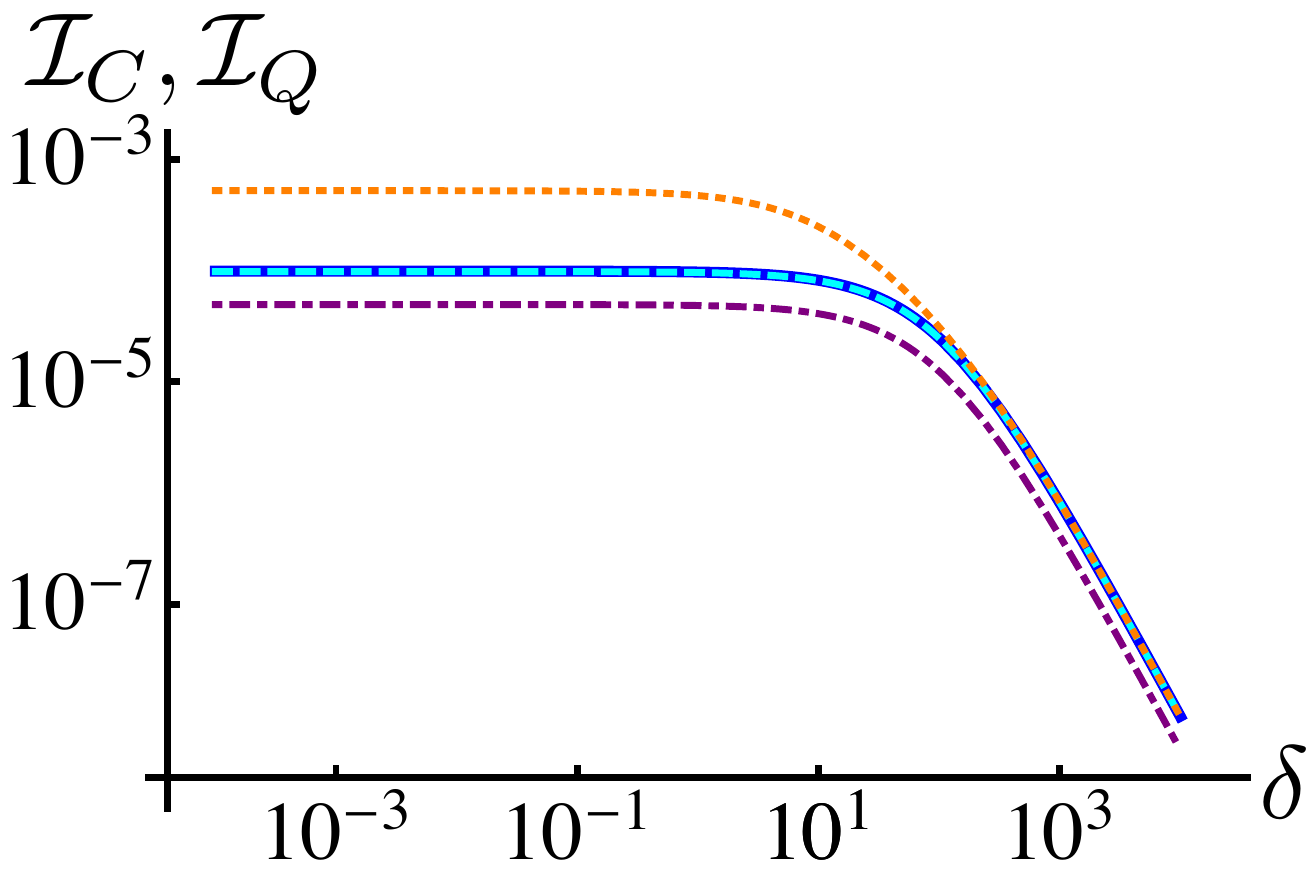}
\caption{(Color online) The QFI for estimation of $\delta$ in the unsqueezed case (blue) and in the squeezed case (orange, dotted) and the Fisher information for a heterodyne measurement in the unsqueezed case (cyan, dashed), and a homodyne measurement in the squeezed case (magenta, dot-dashed) are plotted against $\delta$. Different types of measurement have been chosen to achieve the optimal value of the Fisher information. The oscillator contains $\bar{n}=100$ excitations and is squeezed with squeezing parameter $s = 2.95$ (chosen for convenience of plotting).}
\label{SqueezingPlot}
\end{figure}

We now complement the analysis reported above by assessing whether potential advantages for the estimation performance could arise from the use of a genuinely quantum resource such as squeezing. 
Several different experimental strategies for imposing squeezing on the mechanical oscillator are available~\cite{Agarwal,Kronwald,Vitali,Vanner}. However, rather than proceeding down this path, we qualitatively investigate the effect of introducing squeezing. In the system we describe above, the mechanical subsystem in the steady state is in a slightly squeezed thermal state, and the CSL parameter enters the covariance matrix linearly. We considered a simpler version of this setup: a single oscillator, in a thermal state, characterised by it's initial thermal occupation number $\bar{n}_{th}$. By squeezing the oscillator, we add energy to the system, and the mean occupation number becomes $\bar{n}=\bar{n}_{th}\cosh 2s + \sinh^2 s$, where $s$ is the squeezing parameter. The CSL mechanism, as expressed in Eq.~\eqref{equiv}, appears as an additive contribution to $\bar{n}$. To mimic this, a new ``CSL-type'' parameter $\delta$ is introduced: $\nbar \rightarrow \bar{n}+\delta$. It is this $\delta$ which we estimate. Since this system is also Gaussian, we can use the same procedure as above to calculate $\mathcal{I}_C$ for local Gaussian measurements and $\mathcal{I}_Q$.

Fig.~\ref{SqueezingPlot} shows that estimation of $\delta$ without squeezing yields the same behaviour as shown in Fig.~\ref{FI-QFIoptics} {\bf (a)}, with a flat region followed by a `knee'. In this case the $\mathcal{I}_C=\mathcal{I}_Q$, which we attribute to the measurement being performed directly. When we introduce the squeezing, we see the same behaviour, but with an increase in $\mathcal{I}_Q$, but a complementary decrease in the $\mathcal{I}_C$. So while there is an increase in precision offered by introducing squeezing, this increase is not practically achievable using local Gaussian measurements. We conclude from this qualitative assessment that squeezing would not be a useful resource in the fully-fledged system.



\section{Hybrid optomechanics for discrete-variable probing}\label{s:HybridEst} 
Beside the cavity optomechanical setup addressed so far, where the optical mode serves the purpose of both preparing and probing the mechanical state, we can envision an 
alternative \textit{hybrid scheme}, in which the manipulation is still realized through radiation-pressure interaction with the cavity field but the read-out is carried out via a 
coherent coupling with a two-level system. We assume to have full control on the preparation of the two-level system and to initialize it in the pure state
\begin{equation}
\ket{\psi}=\cos \frac{\vartheta}{2} \ket{0}+
e^{i\varphi}\sin \frac{\vartheta}{2}\ket{1},
\end{equation}
of basis vectors $\{\ket{0},\ket{1}\}$. Here $(\vartheta,\varphi)$ are the angles defining the orientation of the state vector in the Bloch sphere. The Hamiltonian we choose to model 
the interaction between the probe qubit and the mechanical oscillator couples the resonator's position quantum fluctuation to the spin-flip operator, i.e.    
\begin{equation} 
\label{Hint}
\hat H_I= \hbar g\delta\hat{q}\otimes\hat{\sigma}_x \, ,
\end{equation}
with coupling strength $g$.
This interaction model has been derived, as an effective spin-mechanics coupling, from a variety of physical configurations, including a quantum dot grown on mechanical nanostructures~\cite{wilsonrae}, a superconducting qubit coupled to a nanobeam~\cite{rabl}, or a multilevel atom coupled to the field of an optomechanical cavity~\cite{vacanti}. With the exception of the first configuration mentioned here (which would not be suitable for the purposes of our investigation), the motional degrees of freedom of the probe are not involved nor required. We thus assume that the two-level probe is not affected by the CSL mechanism under scrutiny. 
\begin{figure*}[t!]
{\bf (a)}\hskip4cm{\bf (b)}\hskip4cm{\bf (c)}\hskip4cm{\bf (d)}
\includegraphics[width=0.55\columnwidth]{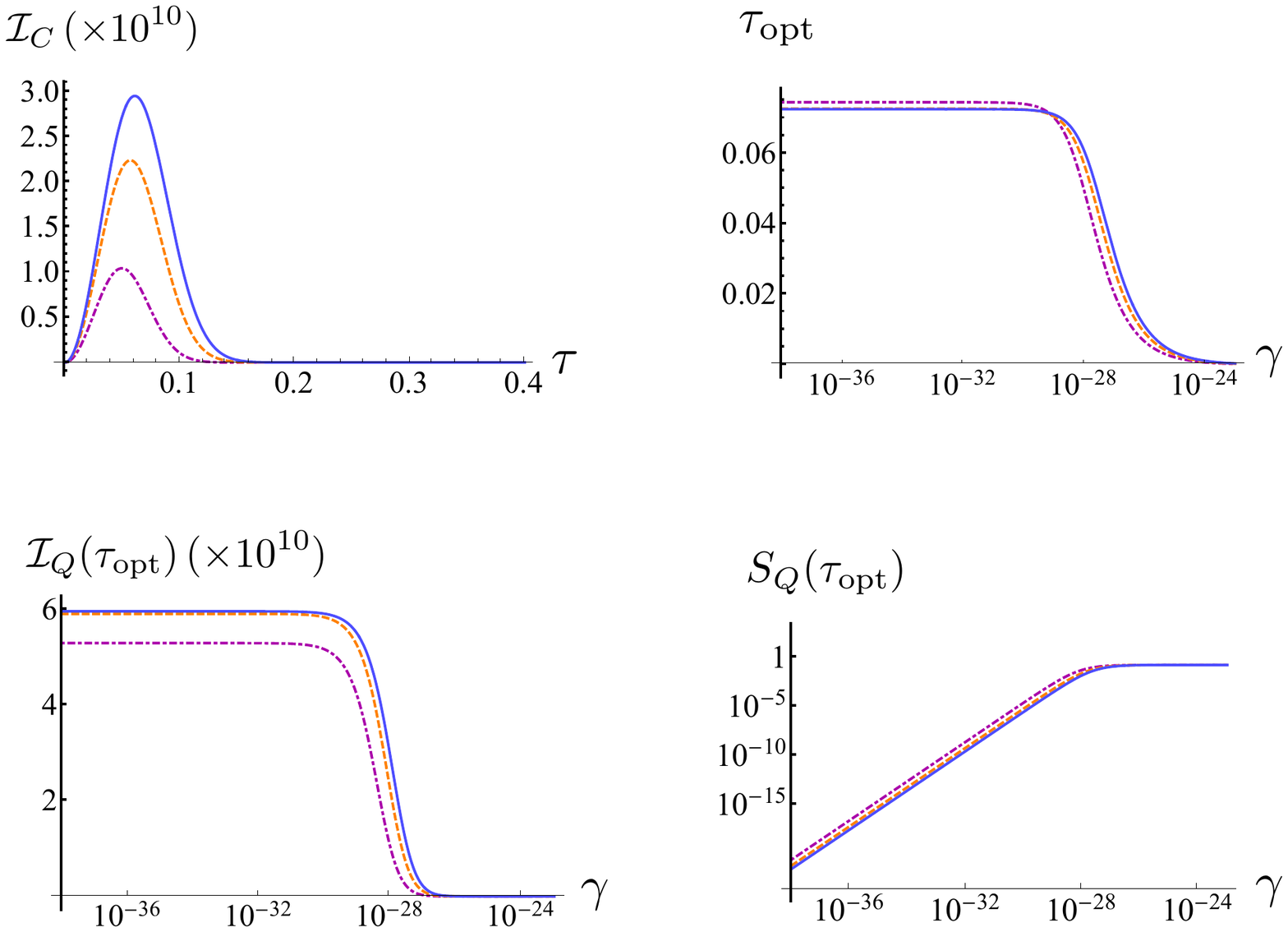}\includegraphics[width=.55\columnwidth]{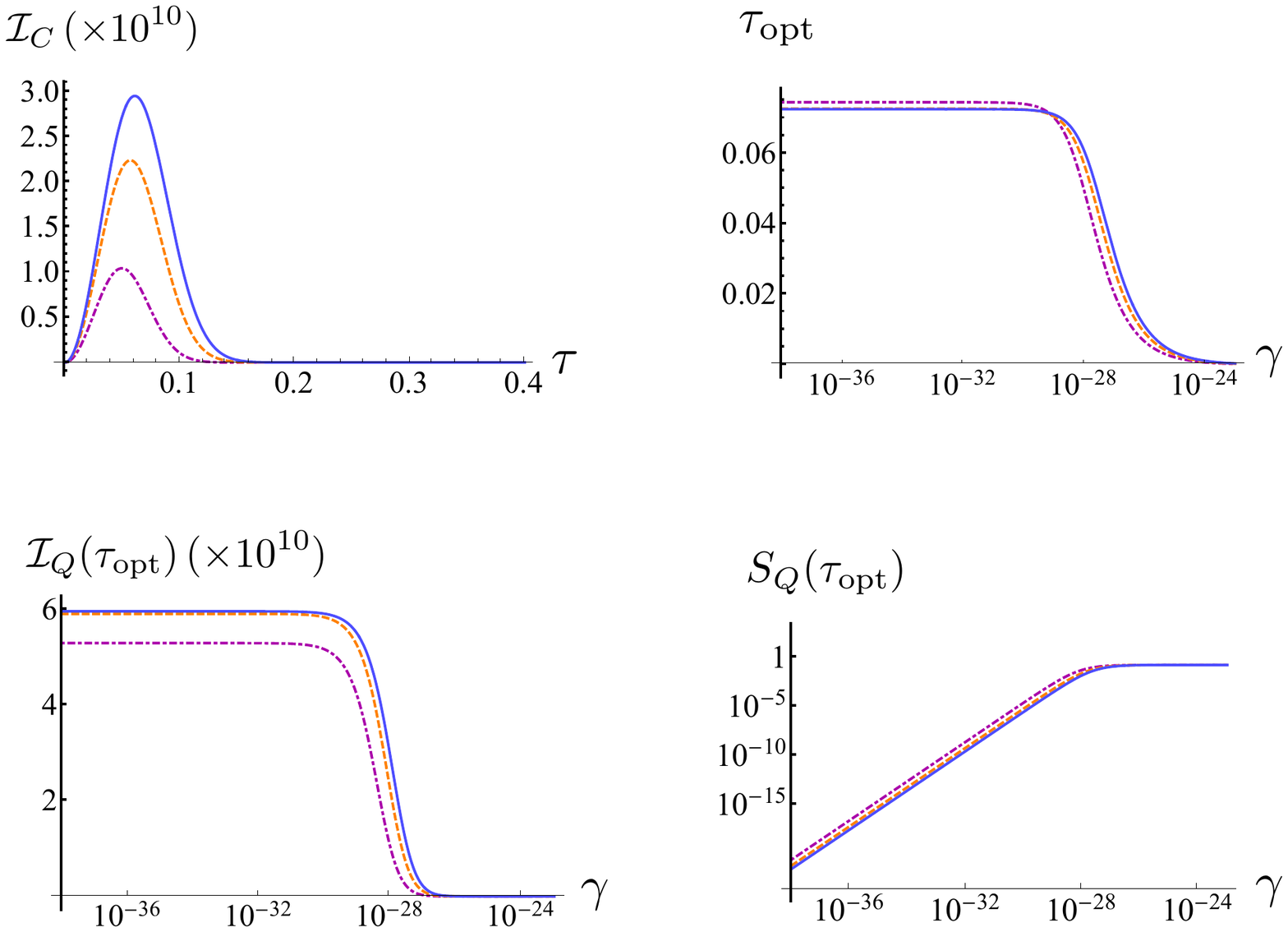}\includegraphics[width=0.55\columnwidth]{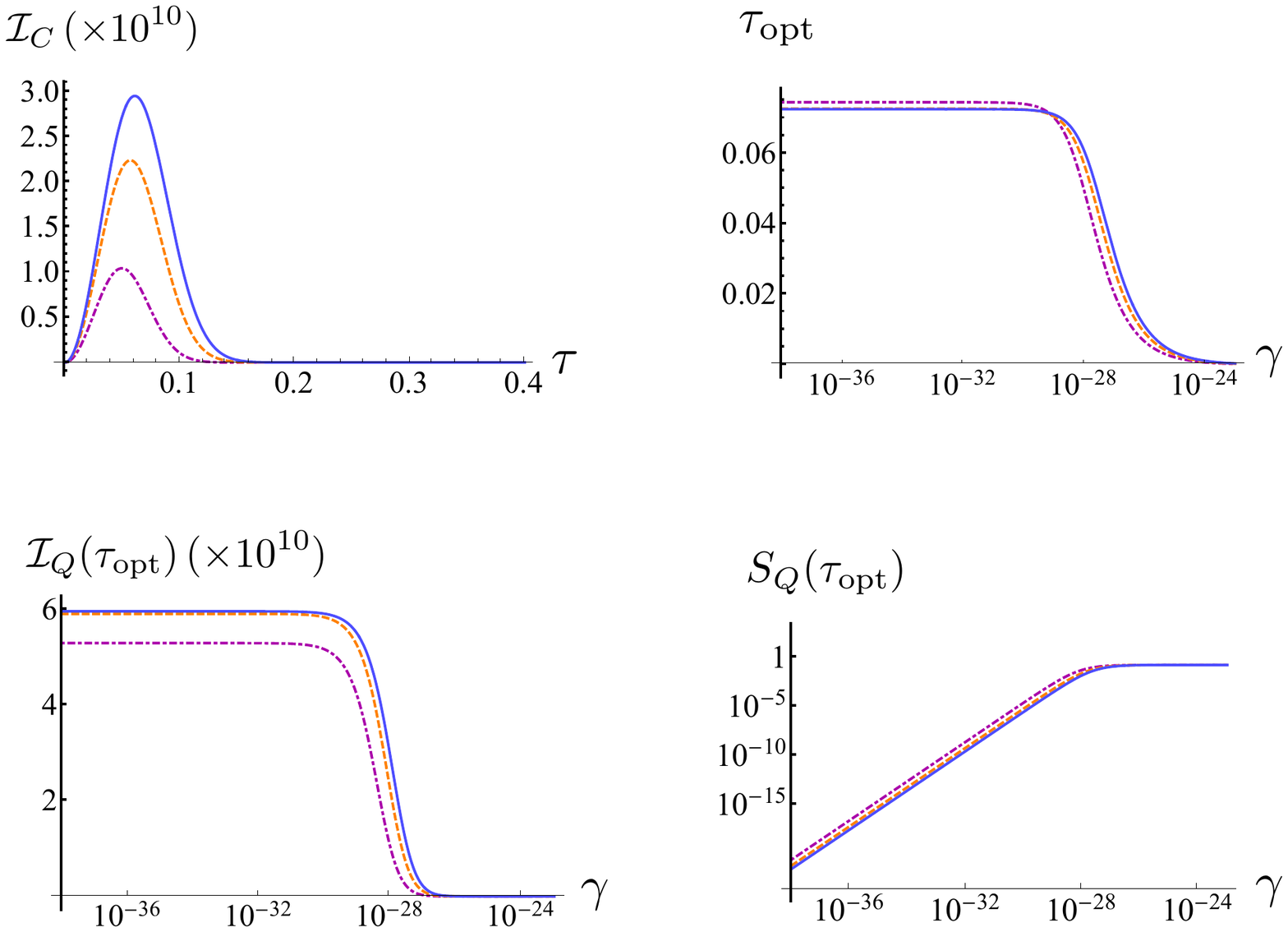}\includegraphics[width=0.55\columnwidth]{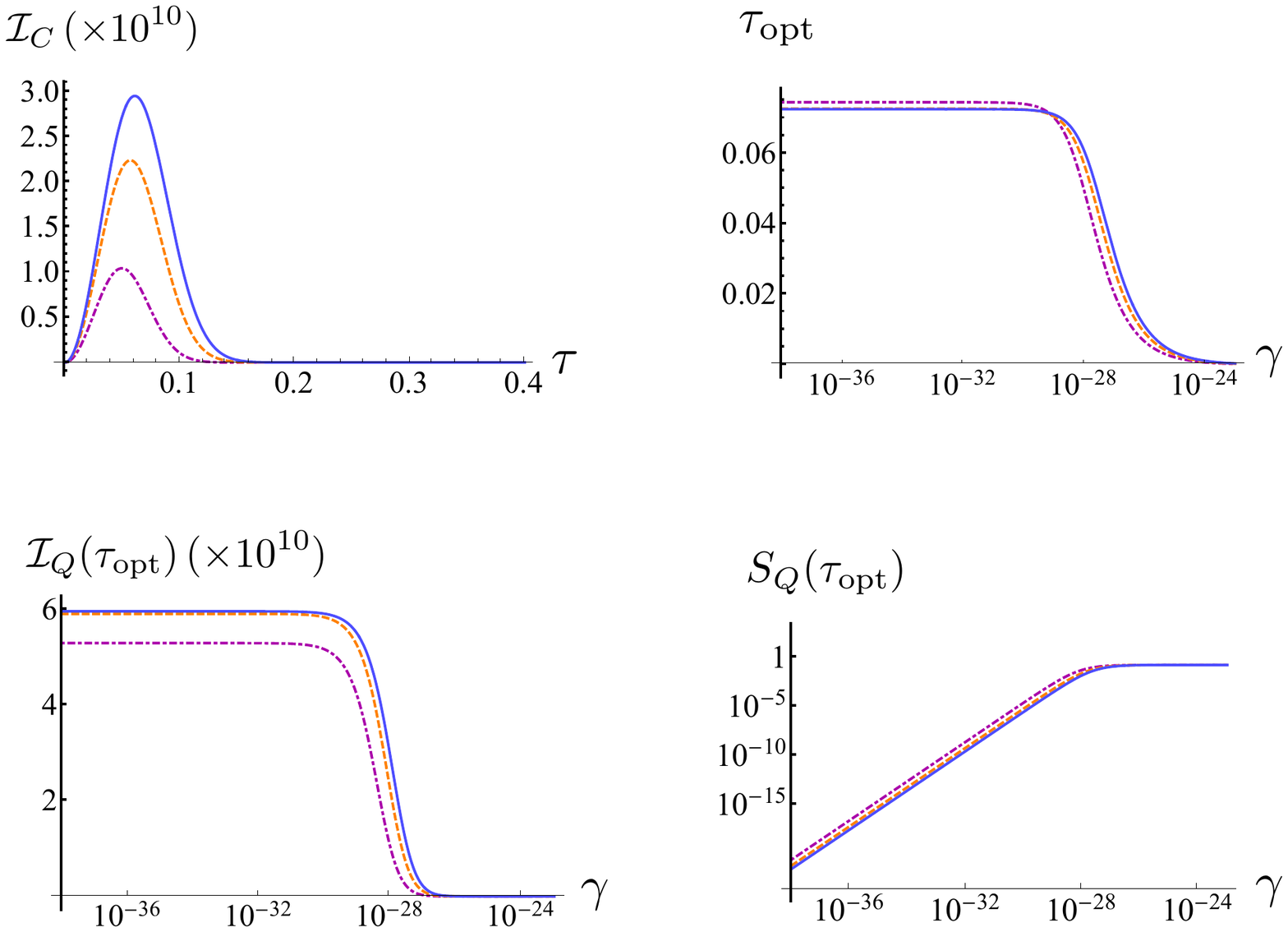}
\caption{(Color online) {\bf (a)} Fisher information against the CSL parameter $\gamma$ for growing values of the effective mass of the mechanical oscillator. We have taken $m=15$ng (magenta dot-dashed curve), $m=150$ng (dashed orange one), and $m=500$ng (solid blue line). We have assumed $\gamma=\gamma_A$. {\bf (b)} For the same values of the mass we study how the optimal measurement time $\tau_{\text{opt}}$ changes with the CSL parameter $\gamma$. {\bf (c)} Plot of the QFI at the optimal time ${\cal I}_Q(\tau_\text{opt})$ and optimised over the probing qubit's preparation plotted against the coupling strength $\gamma$ to the CSL noise field, and relative quantum signal-to-noise ratio $S_Q(\tau_\text{opt})$ {\bf (d)}. Other parameters as in Fig.~\ref{FImauro}.}
\label{FI-QFIhybrid}
\end{figure*}

The qubit-oscillator coupling shown in Eq.~\eqref{Hint} (and its limit under rotating-wave approximation) has already been addressed concerning the estimation of the temperature of 
a mechanical resonator in thermal equilibrium, and the optimality of energy measurements performed on the qubit to this purpose has been shown~\cite{myself1,myself2}. This motivates the choice of that specific form of interaction, given that according to Eq.~\eqref{equiv} the estimation of $\Lambda$ can be mapped to an effective-temperature estimation 
problem. Finally, we assume that no initial correlations are present between the two systems. This can be justified by assuming the optomechanical interaction to be strong enough to quickly prepare the mechanical initial state, which is then coupled to the two-level system through a slow (adiabatic) Hamiltonian~\cite{vacanti,myself2}. 
The measurements are performed on the reduced state of the probe $\hat\varrho_q$  after its joint evolution with the mechanical mode, which is obtained as
\begin{equation}\label{probestate} 
\hat\varrho_q(\tau)= 
\Tr{M}{\hat U_\tau\,\hat\rho_M\otimes|\psi\rangle{}\langle\psi| \, \hat U_{\tau}^{\dagger}},
\end{equation}
where $\hat U_\tau=e^{-i \tau \delta\hat{q}\otimes\hat{\sigma}_x}$ and  $\tau =g t$ is the dimensionless interaction time.
In the Appendix we show that the matrix elements of the probe's state 
\begin{equation}
\hat\varrho_q(\tau)=
\left(\begin{matrix}
\varrho_{00} & \varrho_{01}\\
\varrho_{10} &\varrho_{11}
\end{matrix}\right),\label{varq}
\end{equation}
can be explicitly found to be:
\begin{equation}
\label{probematrix}
\begin{aligned}
\varrho_{00}&=\frac12\bigl(1 + e^{-\zeta}\cos \vartheta \bigr),\quad\varrho_{11}=1-\varrho_{00}, \\
\varrho_{01}&=\varrho^*_{10} =\frac12{\sin \vartheta}\bigl( \cos \varphi -i e^{-\zeta}\sin \varphi \bigr),\\
\end{aligned}
\end{equation}
with $\zeta=2\tau^2\left(\alpha_1+\Lambda \beta_1\right)$. We notice that, since the spin degree of freedom
couples to the resonator's position, only the information about the variance of the mechanical position is copied 
onto the probe. 

\par
The Fisher information Eq. (\ref{FI}) associated with population measurements performed on the probe  is given by
$\mathcal{I}_C(\Lambda)=\sum_{j=0,1}\left[\partial_\Lambda \ln p(j|\Lambda)\right]^2 p(j|\Lambda)$ and takes the 
particularly simple form
\begin{equation}\label{FIhybrid}
\mathcal{I}_C(\Lambda)=\frac{\tau^4 \beta_1^2  \cos^2\vartheta}
{e^{2 \zeta} - \cos^2\vartheta}\,.
\end{equation}
Here, $\mathcal{I}_C$ is a function of the qubit's polar angle $\vartheta$ only, of the interaction time $\tau$, and---through $\alpha_1$ and 
$\beta_1$---of the optical and mechanical parameters in Eq.~\eqref{eq:A}. The parameter of interest $\Lambda$ only 
appears in the exponent $\zeta$. Given the control available over the qubit state preparation, we can maximize the Fisher information 
by choosing $\vartheta=\{0,\pi\}$, i.e. by initializing the qubit either in $\ket{0}$ or $\ket{1}$. In the following an optimal preparation of the
qubit state will be assumed. In Fig.~\ref{FI-QFIhybrid} \textbf{(a)} we show the behavior of the Fisher information as a function of the
interaction time $\tau$, for different values of the resonator's effective mass. As expected, the more massive the oscillating body the more
accuracy gained in the estimation. More importantly, from the picture is apparent the emergence of an optimal probing 
time $\tau_{\text{opt}}$. This optimal time can be evaluated analytically and is given by
\begin{align}
\tau_{\text{opt}}(\Lambda)&=\frac12\sqrt{\left[2+W
\left(-\frac{2}{e^2}\right)\right]/\left(\alpha_1+ \beta_1\Lambda\right)}
\\ & \simeq  0.631 / \sqrt{\alpha_1+ \beta_1\Lambda}\notag
\end{align}
with $W(y)$ the Lambert function of argument $y$~\cite{corless1996}.
The behavior of $\tau_{\text{opt}}$ as a function of $\gamma$ is shown in  Fig.~\ref{FI-QFIhybrid} \textbf{(b)}. 
Remarkably, $\tau_{\text{opt}}$ exhibits a sensitive variation in the region $\gamma\approx \gamma_A$ while for smaller values the optimal interaction time becomes
almost independent on $\gamma$, so that no fine tuning of the measurement time would be needed.   
Finally, since according to Eq.~\eqref{FIhybrid} the Fisher information is proportional to $\beta_1^2$, we can look closer to this term: By expanding $\beta_1$ into power 
of $\chi/\omega_m$, the leading term is independent on the optomechanical coupling and reads $(2 \gamma_m)^{-1}$. Therefore, a reduction of the mechanical losses would lead to better
performances in the estimation of the strength of the collapse mechanism. This could be expected, and it is also in agreement with Eq.~\eqref{equiv}.  
\par
In order to evaluate the QFI, we diagonalize the state of the probe as $\varrho_q=\varrho_+\ket{\psi_+}\bra{\psi_+}+\varrho_-\ket{\psi_-}\bra{\psi_-}$,
which enables to cast Eq.~\eqref{QFI} in the explicit form
\begin{align}
\mathcal{I}_Q(\Lambda)=&\sum_{k=\pm}\frac{(\partial_\Lambda \varrho_k)^2}{\varrho_k}
+2 c \sum_{k\neq l=\pm}\bigl|
\sum_{j=0,1}(\partial_\Lambda\langle{j}|\psi_{k}\rangle)\langle\psi_l|j\rangle
\bigr|^2 
\end{align}
with $c=(1-2\varrho_{+})^2$.
The actual calculation, which produces expressions too involved to be
reported here, shows that $\mathcal{I}_Q(\Lambda)$ depends on both the angles entering the initial qubit's state and 
is maximized for two independent sets of choices of the qubit-state parameters: One can either prepare
the qubit in one of the basis states $\ket{0}$ or $\ket{1}$, or choose $\varphi=\{  \pi/2,
3\pi/2 \}$, regardless of $\vartheta$. In  Fig.~\ref{FI-QFIhybrid} \textbf{(c)} we show the QFI
for optimal state preparation evaluated at the optimal time $\tau_{\text{opt}}$ against $\gamma$, for increasing values
of the resonator's mass, while in panel \textbf{(d)} the corresponding quantum signal-to-noise ratio $S_Q(\tau_{\text{opt}})$ is shown. 
\par 
It is remarkable that the behavior of ${\cal I}_Q(\tau_\text{opt})$ and $S_Q(\tau_{\text{opt}})$ against $\gamma$ is very similar to  the trends shown in Figs.~\ref{FImauro} and \ref{FI-QFIoptics}, notwithstanding the significant differences between the measurement strategies being pursued, which is indicative of a profound fundamental reason behind the insensitivity of the estimation performance for values of $\gamma$ smaller than the value inferred by Adler and, on the other hand, the quick depletion of the estimation precision for $\gamma\ge\gamma_A$.
 
\noindent
\section{Conclusions}\label{s:Conclusions} 
We have assessed the important collapse model provided by the CSL mechanism from the perspective of quantum estimation theory. We have provided key information on the actual experimental approach towards the estimation of important features of the model through state-of-the-art methods in cavity optomechanics. Moreover, our investigation allowed us to pinpoint important qualitative and quantitative differences in the estimation accuracy of the strength of the coupling between a mechanical oscillator and the noise field responsible for the CSL reduction as its value is  varied within the range of values currently considered as plausible. 

Our study bridges cavity optomechanics, fundamental collapse-model theory, and sophisticated inference techniques borrowed from quantum information theory towards the systematic assessment of fundamental reduction models. 

During the completion of this paper, we became aware of the related work by M. G. Genoni, {\it et al.}, arXiv:1605.09168, which addresses the discrimination of reduction models through continuous-time measurements of mechanical oscillators. 

\noindent
\section*{Acknowledgements}
MB and MP are grateful to Marco Genoni and Alessio Serafini for discussions. We acknowledge support from the EU projects TherMiQ and QuProCS, the Universita' degli Studi di Milano through the H2020 Transition Grant 15-6-3008000-625, the John Templeton Foundation (grant numbers 39530 and 43467), the Julian Schwinger Foundation (grant number JSF-14-7-0000), the UK EPSRC (grants EP/M003019/1 and EP/J014664/1 and the EPSRC DTA to Queen's University Belfast), the Leverhulme Trust (RPG-2016-046), and the Foundational Questions Institute [FQXi].

\section*{Appendix}

\subsection*{Appendix A:  Calculation of the QFI} 
In order to find $\Phi$, it is necessary to perform a symplectic diagonalisation of the equation $\partial_\Lambda{\sigma} =2 \sigma\Omega \Phi \Omega^T \sigma - \frac{\Phi}{2}$. Using Williamson theorem, we define 
$S\sigma S^T = \sigma_S = \textrm{Diag}(d_1,d_1,\dots,d_n,d_n)$,
where $S$ is a suitable symplectic matrix, i.e. a matrix such that $S\Omega S^T=\Omega$. This leads to
\begin{equation}
(\Phi_S)_{ij}=\frac{\left(\Omega^T \sigma_S\partial_\Lambda{\sigma}_S\sigma_S\Omega + \frac{1}{4}\partial_\Lambda{\sigma}_S\right)_{ij}}{2 d^2_i d^2_j - \frac{1}{8}}
\end{equation}
where $\Phi_S=S\Phi S^T$, $\partial_\Lambda{\sigma}_S=S\partial_\Lambda{\sigma}S^T$, and $d_j$ are the symplectic eigenvalues of $\sigma$, with $d_j=d_{j-n}$ for $j>n$. We can then obtain $\Phi$ from an inverse symplectic transformation, to then find the QFI 
\begin{equation}\label{QFIapndx}
\mathcal{I}_Q(\Lambda) = \textrm{tr}\left[\Omega^T(\partial_\Lambda{\sigma})\Omega\Phi\right] + (\partial_\Lambda{\bm \xi}^T)\sigma^{-1}(\partial_\Lambda{\bm \xi}).
\end{equation}
This equation holds for general (pure or mixed) $n$-mode Gaussian states, however, since we measured the light mode to find the Fisher information, we are interested in the single mode case.  The symplectic diagonalisation therefore has the form $\sigma_S =\sqrt{\det{\sigma}}\openone$ (with $\openone$ the $2\times2$ identity matrix), and $S=(\det{\sigma})^{\frac{1}{4}}/\sqrt{\sigma}$, and Eq.~(\ref{QFIapndx}) simplifies to Eq.~(\ref{QFI2}).

\subsection*{Appendix B: Hybrid spin-optomechanics calculation of the FI}
Here we derive the expression of the probe's matrix elements appearing in Eq.~\eqref{probematrix}. 
If we express the evolution due to the spin-boson coupling Eq.~\eqref{Hint} in terms of the $\sigma_x$-eigenstates $\ket{\pm}=(\ket{0}\pm\ket{1})/\sqrt{2}$ we get
$\hat U_\tau=e^{-i \tau \delta\hat{q}\otimes\hat{\sigma}_x}=D\bigl(- \frac{i\tau}{\sqrt{2}}\bigr)\otimes \ket{+}\bra{+}+D\bigl( \frac{i\tau}{\sqrt{2}}\bigr)\otimes \ket{-}\bra{-}$
and and $\tau =g t$. That is, in that basis the evolution is just a conditional displacement of the mechanical state $\varrho_M$. The latter can be suitably decomposed
in the phase space spanned by the two eigenvalues $(\delta q, \delta p)$ of the mechanical fluctuation operators as   
\begin{equation}
\varrho_M=\int_{\mathbb{C}} \mathrm{d}^2\xi \,\chi_M(\xi) \hat D^{\dagger}(\xi) \, ,
\end{equation}
where $\mathrm{d}^2\xi=\mathrm{d}\xi_r \mathrm{d}\xi_i$, $\hat D(\xi) = \exp\{i\sqrt2 \xi_i \delta\hat{q}-i\sqrt2 \xi_r \delta\hat{p} \}$ is the
displacement operator and $\chi_M(\xi)=\Tr{}{\varrho_M \hat D(\xi)}$ is the characteristic function of the mechanical state. The calculation of the reduced state Eq.~\eqref{probestate} 
is then naturally carried out in the $\{\ket{+},\ket{-}\}$ basis, i.e.
\begin{equation}
\varrho_q(\tau)=
\left(\begin{matrix}
\varrho_{++} & \varrho_{+-}\\
\varrho_{-+} &\varrho_{--}
\end{matrix}\right)\, ,\label{varq}
\end{equation}
where the matrix elements read
\begin{align}
\varrho_{++}&=\frac12(1+\sin \theta \cos \phi) \,,\notag \\ 
\varrho_{+-}&=\varrho_{-+}^*=\frac12(\cos \theta + i \sin \theta \sin \phi) \chi_M(-i\sqrt2 \tau) \,,\notag \\
\varrho_{--}&=1-\varrho_{++}\, . \notag
\end{align}
In order to evaluate the contribution of the characteristic function $\chi_M(-i\sqrt2 \tau)$ 
we start from the Wigner function of the mechanical state 
\begin{equation}
W(\alpha)=\frac{1}{\pi \sqrt{\det\sigma_M}}e^{-\frac12(\sqrt2\alpha_r,\sqrt2\alpha_i)^T\sigma_M^{-1}(\sqrt2\alpha_r,\sqrt2\alpha_i)} \, ,
\end{equation}
for which we have the covariance matrix $\sigma_M=\mathrm{diag}(\alpha_1+\Lambda \beta_1,\alpha_2+\Lambda \beta_2)$, and
take its complex Fourier transform $\chi_M(\xi)=\int \mathrm{d}^2\alpha W(\alpha) e^{\alpha^*\xi-\alpha \xi^*}$. Being a 
Gaussian integral, it is easily performed and we find $\chi_M(-i\sqrt2 \tau)=e^{-2\tau^2(\alpha_1+\Lambda \beta_1)}$. 
Moving back to the computational basis we finally find the expressions appearing in Eq.~\eqref{probematrix}.

\subsection*{Appendix C: Mechanical covariance matrix elements}\label{s:MatrixElements}
The explicit form of $\alpha_{1,2}$ and $\beta_{1,2}$ from the mechanical covariance matrix in Eq.~\eqref{mechcovmtrx} is as follows
\begin{widetext}
\begin{equation}
\begin{aligned}
\alpha_1 &= \frac{1}{A}[\omega _m  (2 \kappa  \omega _m^2  (\Delta ^2+\kappa ^2 ) (\kappa  \chi ^2+\gamma _m  (2 \gamma _m (\kappa +2 \kappa   \nbar)-2 (2  \nbar+1) (\Delta -\kappa ) (\Delta +\kappa )+\chi ^2 ) )\\
&+\Delta  \chi ^2 \omega _m  (-2 \kappa ^2 \chi ^2+\gamma _m^2(2  \nbar+1)  (\Delta ^2-3 \kappa ^2 ) -\kappa  \gamma _m  (\chi ^2-4 (2  \nbar+1) (\Delta -\kappa ) (\Delta +\kappa ) ) )\\
&+2 \kappa   (\Delta ^2+\kappa ^2 )  (\Delta ^2+\kappa ^2+2 \kappa  \gamma _m+\gamma _m^2 )  (\kappa  \chi ^2+ \gamma _m(2  \nbar+1)  (\Delta ^2+\kappa ^2 ) )\\
&+2 \kappa \gamma _m \omega _m^4 (2  \nbar+1)  (\Delta ^2+\kappa ^2 ) -2 \Delta  \kappa \chi ^2 \gamma _m \omega _m^3 (2  \nbar+1) )],\\
\beta_1 &= \frac{1}{A}[\omega _m  (\Delta  \chi ^2 \omega _m  (4 \kappa  (\Delta -\kappa ) (\Delta +\kappa )+  \gamma _m (\Delta ^2-3 \kappa ^2 ))+4 \kappa \omega _m^2   (\Delta ^2+\kappa ^2 ) (-\Delta ^2+\kappa ^2+\kappa  \gamma _m )\\
&+2 \kappa^2 \omega _m^4  (\Delta ^2+\kappa ^2 )^2  (\Delta ^2+\kappa ^2+2 \kappa  \gamma _m+\gamma _m^2 )  (\Delta ^2+\kappa ^2 ) -2 \Delta  \kappa  \chi ^2 \omega _m^3 )],\\
\alpha_2  & = \frac{1}{B}[2 \kappa  \omega _m^2 \gamma _m (\kappa  \chi ^2+2 (2  \nbar+1)  (-\Delta ^2+\kappa ^2+\kappa  \gamma _m ) )+2 \kappa   (\Delta ^2+\kappa ^2 )  (\kappa  \chi ^2+\gamma _m  (\gamma _m(2  \nbar+1)  (2 \kappa +\gamma _m )+\\
&(2  \nbar+1)  (\Delta ^2+\kappa ^2 )+\chi ^2 ) )+\Delta \chi ^2 \gamma _m \omega _m (2  \nbar+1)  (2 \kappa +\gamma _m )+2 \kappa \gamma _m \omega _m^4 (2  \nbar+1) ],\\
\beta_2  &= \frac{1}{B}[4 \kappa  \omega _m^2  (-\Delta ^2+\kappa ^2+\kappa  \gamma _m )+2 \kappa   (\Delta ^2+\kappa ^2 )  (\Delta ^2+\kappa ^2+2 \kappa  \gamma _m+\gamma _m^2 )+\Delta  \chi ^2 \omega _m  (2 \kappa +\gamma _m )+2 \kappa  \omega _m^4],\\
&\textrm{where the denominators $A$ and $B$ are given by}\\
A & = 2  ( \omega _m(\Delta ^2+\kappa ^2 ) -\Delta  \chi ^2 )  (4 \kappa  \gamma _m \omega _m^2  (-\Delta ^2+\kappa ^2+\kappa  \gamma _m )+2 \kappa  \gamma _m   (\Delta ^2+\kappa ^2 ) (\Delta ^2+\kappa ^2+2 \kappa  \gamma _m+\gamma _m^2 )\\
&+\Delta  \chi ^2 \omega _m  (2 \kappa +\gamma _m ){}^2+2 \kappa  \gamma _m \omega _m^4 ),\\
B &= 8 \kappa  \gamma _m \omega _m^2  (-\Delta ^2+\kappa ^2+\kappa  \gamma _m )+4 \kappa  \gamma _m   (\Delta ^2+\kappa ^2 ) (\Delta ^2+\kappa ^2+2 \kappa  \gamma _m+\gamma _m^2 )+2 \Delta  \chi ^2 \omega _m  (2 \kappa +\gamma _m ){}^2+4 \kappa  \gamma _m \omega _m^4.
\end{aligned}
\end{equation}
\end{widetext}

\end{document}